\documentclass[aps,pra,twocolumn,showpacs]{revtex4}

\usepackage{graphicx}
\usepackage{color}
\usepackage{bm}
\usepackage{amsmath, amsthm, amssymb,mathrsfs}
\usepackage{csquotes}
\usepackage[colorlinks=true, citecolor=blue,allcolors=blue]{hyperref}
\newcommand{\Schro}{Schr\"odinger}
\newcommand{\MESA}{{\footnotesize\textsc{MESA}}}
\newcommand{\STOCK}{{\footnotesize\textsc{STOCK}}}

\newcommand{\braket}[2]{\langle#1|#2\rangle}
\newcommand{\braOket}[3]{\langle#1|#2|#3\rangle}
\DeclareMathOperator*{\SumInt}{%
\mathchoice%
  {\ooalign{$\displaystyle\sum$\cr\hidewidth$\displaystyle\int$\hidewidth\cr}}
  {\ooalign{\raisebox{.14\height}{\scalebox{.7}{$\textstyle\sum$}}\cr\hidewidth$\textstyle\int$\hidewidth\cr}}
  {\ooalign{\raisebox{.2\height}{\scalebox{.6}{$\scriptstyle\sum$}}\cr$\scriptstyle\int$\cr}}
  {\ooalign{\raisebox{.2\height}{\scalebox{.6}{$\scriptstyle\sum$}}\cr$\scriptstyle\int$\cr}}}

\begin{document}

\title{Application of the Complex Kohn Variational Method to Attosecond Spectroscopy}

\author{N. Douguet$^1$, B.~I.~Schneider$^2$, and L. Argenti$^1$}

\affiliation{$^1$Department of Physics, University of Central Florida, Orlando, Florida 32186, USA}
\affiliation{$^2$Physics Division, National Science Foundation, Gaithersburg, MD 20899, US}

\date{\today}


\begin{abstract}
 The complex Kohn variational method is extended to compute light-driven electronic transitions between continuum wavefunctions in atomic and molecular systems. This development enables the study of multiphoton processes in the perturbative regime for arbitrary light polarization. As a proof of principle, we apply the method to compute the photoelectron spectrum arising from the pump-probe two-photon ionization of helium induced by a sequence of extreme ultraviolet and infrared-light pulses. We compare several two-photon ionization pump-probe spectra, resonant with the ($2s2p$)$^1P^o_1$ Feshbach resonance, with independent simulations based on the {\it atomic} B-spline close-coupling STOCK code, and find good agreement between the two approaches. This new finite-pulse perturbative approach is a step towards the \emph{ab initio} study of weak-field attosecond processes in poly-electronic molecules.

\end{abstract}

\maketitle

\section{INTRODUCTION}
The availability of attosecond sources \cite{Krausz09} covering a wide frequency domain~\cite{Chini2013,Liao17,10.1088/2040-8986/aaa394, Popmintchev30062009,Xiong2009,Ishii2014,Johnson2016,Teichmann2016,PhysRevLett.105.173901,Li2016,Stein2016,Popmintchev2012} has allowed experimentalists to explore ultrafast processes in gases \cite{Chini2014} and solids \cite{Lepine2014,Schultze2014}. Experimental pump-probe techniques with subfemtosecond time-resolution~\cite{PhysRevLett.105.263002,PhysRevLett.108.233002}, such as attosecond transient absorption spectroscopy (ATAS) \cite{PhysRevA.86.063411,PhysRevLett.106.123601,PhysRevLett.109.073601,Chini2013,Liao17}, and RABBITT (reconstruction of attosecond beating by interference of two-photon transitions), a photoelectron interferometric technique~\cite{Heuser2015,Gruson734,Beaulieu2017},  are key to investigate ultra-fast dynamics in molecules and atoms. In particular, attosecond spectroscopy allows us to highlight and resolve in time autoionization decay, which is driven by electron correlation, the role of coherent valence excitation in dissociating molecules, and the emergence of circular dichroism in the photoemission from chiral systems~\cite{doi:10.1021/acs.chemrev.6b00453}.
For example, the complete characterization of an electron wave packet created through an autoionizing resonance has recently been achieved in helium \cite{Gruson734}, oxygen autoionizing Rydberg states were studied in an electronically- and vibrationally-resolved fashion with ATAS~\cite{Liao17}, and the dynamics of systems as complex as CF$_4^+$ and SF$_6^+$ was studied through time-resolved x-ray absorption spectroscopy~\cite{Pertot2017}. An attosecond photoelectron interferometry study on camphor enantiomers~\cite{Beaulieu2017} revealed an angular dependence of the photoemission delay as large as 24 attoseconds (as), between electrons ejected forward and backward upon photoionization by circularly polarized light.

The development of accurate computational and theoretical tools to treat multi-photon processes in complex molecules is essential to the interpretation of experimental results in attosecond dynamics~\cite{Palacios3973,PhysRevA.91.012509,doi:10.1021/acs.chemrev.6b00453,Klinder18}. The computational treatment of such processes is challenging as it requires an explicit representation of multi-electronic bound and continuum states. In particular, crucial aspects of atomic and molecular ionization, such as autoionizing resonances, and inter-channel coupling, cannot be satisfactorily accounted for within  single-active-electron approaches.

In recent years, our understanding of multiphoton processes in correlated systems has made progress using either direct integration of the time-dependent {\Schro} equation or perturbative calculations, and developing new non-Gaussian or hybrid bases for the expansion of the electronic wavefunction~\cite{PhysRevA.96.022507,PhysRevA.91.012509,Palacios3973,Marante17,PhysRevA.90.012506,Klinder18,C6FD00093B,PhysRevA.96.053421,doi:10.1021/acs.chemrev.6b00453}. Here, we present a novel adaptation of the complex-Kohn (CK) variational method~\cite{Rescigno95a,Rescigno95b} to compute free-free electronic transition dipole moments in polyelectronic systems, with applications to multiphoton processes in the perturbative regime.

The present implementation of the CK method~\cite{Schneider90,Rescigno02} combines the \MESA~\cite{mesa90} quantum chemistry package with Coulomb functions and an adaptive-grid method \cite{McCurdy89} to compute the bound-bound, bound-free, and free-free integrals that are needed to evaluate the elements of the electronic Hamiltonian within the close-coupling fixed-nuclei approximation. The transition dipole matrix elements from a bound state to the multi-channel continuum states are variationally optimal~\cite{PhysRevA.41.1695,Spruch72}. Over the past three decades, the CK method has proven to accurately represent the molecular electronic continuum as demonstrated by its successful description of processes such as dissociative recombination~\cite{Chourou11,Douguet12,Fonseca14,Larson17}, dissociative electron attachment~\cite{Slaughter13,Douguet15}, vibrational excitation~\cite{Schneider90,Rescigno02}, photoionization~\cite{Rescigno93,Williams12,Jose14,Douguet13,McCurdy17}, and photodetachment~\cite{Miyabe11,Douguet14}. 

The CK method has also been successfully employed to compute the photoionization cross sections of mid-sized molecules such as methanol~\cite{Slaughter13}, SF$_6$~\cite{Jose14} or CF$_4$~\cite{McCurdy17}. In addition, the CK code has the option of using effective core potentials~\cite{Pacios85,Rescigno96}, and hence it can be applied to larger systems as long as the interaction between core electrons and external time-dependent fields can be neglected.
 
Bound-free and free-free electronic transition dipole moments can be used in lowest-order perturbation theory (LOPT) to treat multiphoton processes in the weak-field regime~\cite{Gryzlova18,PhysRevA.93.023429}. When applicable, perturbative approaches offer two key advantages over time-step integration of the time-dependent Schr{\"o}dinger equation (TDSE). First, they are efficient, as they can predict with a modest computational cost the outcome of the interaction with an arbitrary (and hence arbitrarily long) sequence of pulses and for any given polarization. Second, they offer insight on a reaction mechanism, as they allow one to disentangle the contribution of individual quantum paths to a given process.

The application of perturbative methods in the multiphoton ionization of complex systems has mostly been limited by the need of computing free-free electronic transition dipole moments, which are notoriously challenging to estimate, as evidenced by several theoretical studies~\cite{Madajczyk89,Veniard90,Korol93,Mercouris94,Nikolopoulos06,Novikov11,Komninos12}. On the other hand, several accurate methods, such as the $R$-matrix theory \cite{R_Matrix}, the Schwinger variational method \cite{Lucchese1983}, the complex Kohn variational method \cite{Rescigno95a,Rescigno95b}, and the XCHEM method \cite{Klinder18} can now provide accurate multi-electron continuum wavefunctions and could be employed to evaluate free-free transition dipole moments in multi-electronic systems. To our knowledge, none of these methods has yet been used to this end.

In this work, we consider two-photon ionization of the $1s^2$~{$^1$S$^e$} ground state of helium as a proof of principle of the method on a simple atomic target and using a relatively small calculation. Two-photon ionization of helium is induced by overlapping extreme ultraviolet (XUV) with either infrared (IR) or visible (VIS) pulses, with the XUV tuned on the $2s2p$~{$^1$P$^o$} Feshbach resonance. Even though the method can treat pulses with arbitrary light polarization (e.g. orthogonal or elliptically polarized), here we consider  linearly polarized XUV and IR pulses along the same axis to simplify the presentation of the approach. The free-free dipole moments are compared with results obtained from an independent stationary calculation performed with an extension of the \STOCK~B-spline close-coupling code for atomic ionization~\cite{STOCK2013}. Furthermore, the photoelectron spectra obtained with the present finite-pulse perturbative method are compared with those computed by integrating numerically the time-dependent Schr\"{o}dinger equation in the atomic  \STOCK~close-coupling basis. The current calculations have reproduced a pump-probe experiment in which two consecutive odd high-harmonics of a fundamental visible pump, are followed by a weak replica acting as a probe. This pump-probe scheme, which is at the basis of RABBITT spectroscopy~\cite{Beaulieu2017,Paul1689,Gruson734,PhysRevLett.113.263001}, creates multiple two-photon
pathways to the harmonic sidebands whose interference as a function of the pump-probe delay enables one to retrieve the phase of the resonant ionization amplitude.
In all these comparisons we find good agreement between the present method and independent benchmark simulations, which indicate that stationary continuum calculations can be fruitfully used in association with finite-pulse perturbative formulas to predict the outcome of pump-probe simulations in the weak-field limit.

The paper is organized as follows. In Sec.~\ref{sec:1} we summarize the CK method, its extension to compute the free-free transition dipole moments, and the perturbative approach used to evaluate stationary and finite-pulse two-photon amplitudes. Our results are presented and discussed in Sec. \ref{sec:2}. Finally, in Sec. \ref{sec:3} we offer our conclusions and outlook.  Unless stated otherwise, atomic units are used throughout.

\section{The theoretical approach}
\label{sec:1}
\subsection{The complex Kohn wavefunction}
\label{sec:1}
In the CK variational method, the $(N+1)$-electron non-relativistic wavefunction is expanded in the close-coupling form
\begin{eqnarray}
\label{eq:MESA_CCExpansion}
\psi_{\Gamma E} &&= \sum_{\Gamma'}\hat{\mathcal{A}} [\Phi_{\Gamma'}(\mathbf{x}_1,\cdots,\mathbf{x}_{N},\zeta_{N+1})F_{\Gamma',\Gamma E}(\mathbf{r}_{N+1})]  \nonumber\\
&&+\sum_{\mu}d_{\mu}^{\Gamma E}\Theta_\mu(\mathbf{x}_1,\cdots,\mathbf{x}_{N+1}),
\end{eqnarray}
where $\mathbf{x}_{i}=(\mathbf{r}_i,\zeta_i)$ are the position and the spin variables of the $i-$th electron and $\Phi_{\Gamma}(\mathbf{x}_1\cdots\mathbf{x}_{N},\zeta_{N+1})$ are the channel functions for a parent-ion state.  The channel functions are built from a set of {\it internal} orbitals, coupled to the photoelectron spin $\zeta_{N+1}$.  This leads to a well defined total multiplicity $2S+1$ and spin projection $\Sigma$ for the wavefunction. The projector \hbox{$\hat{\mathcal{A}}$} ($\hat{\mathcal{A}}^2=\hat{\mathcal{A}}$) ensures the wave function is antisymmetric for all electrons. The channel label $\Gamma$ specifies the state of the ion, the total multiplicity, and the asymptotic photoelectron angular quantum numbers $\ell$ and $m$. The close-coupling expansion over $\Gamma'$ typically runs over both open and closed channels at the total electronic energy $E$ of interest.  The internal orbitals are chosen as a subset of the total set of molecular orbitals used in the calculation and the configuration state functions (CSFs) used to construct $\Phi_{\Gamma}(\mathbf{x}_1\cdots\mathbf{x}_{N},\zeta_{N+1})$ are built from these internal orbitals and are typically designed to yield reasonably accurate representations of the ground and excited ionic states of the target.    The elements $\Theta_\mu$ are $(N+1)$ electron CSFs built exclusively from  {\it internal orbitals} and are employed to describe short range electron correlation effects. Henceforth, we refer to the space spanned by the internal orbitals as the {\it reference} space. 

In the case of open channels, the scattering orbitals $F_{\Gamma',\Gamma E}(r)$ are expanded in a set of {\it external} molecular orbitals $\varphi_\sigma$ and continuum functions $f^{\Gamma E}_{\ell m}(\mathbf{r})$ and $h^{\Gamma' E}_{\ell'm'}(\mathbf{r})$, 
\begin{eqnarray}
\label{eq:MESA_ScatteringExpansion}
F_{\Gamma',\Gamma E}(\mathbf{r}) &=& \sum_{\sigma}c_\sigma^{\Gamma',\Gamma E}\varphi_{\sigma}(\mathbf{r}) \nonumber\\
&+&\sum_{\ell' m'} [f^{\Gamma E}_{\ell m}(\mathbf{r})\delta_{\Gamma'\Gamma} +
h^{\Gamma' E}_{\ell'm'}(\mathbf{r})T^{\Gamma'\Gamma E}_{\ell' m',\ell m}].
\end{eqnarray}
The continuum functions $f^{\Gamma E}_{\ell m}$ and $h^{\Gamma E}_{\ell m}$ asymptotically approach the regular and irregular-outgoing Coulomb function, respectively, with energy $\epsilon_{\Gamma'}=E-E_{\Gamma'}$ and orbital angular momentum $\ell$, 
\begin{eqnarray}
\label{eq:Coulomb_functions}
 f^{\Gamma E}_{\ell m}(\mathbf{r}) &\sim& \frac{\sin \theta_\Gamma (r)}{\sqrt{k_\Gamma} r} Y_{\ell m}(\hat{r}) ,\\
h^{\Gamma E}_{\ell m}(\mathbf{r})& \sim& \frac{\exp[ i\theta_\Gamma (r)]}{\sqrt{k_\Gamma} r} Y_{\ell m}(\hat{r}),\label{eq:outgoing}\\
\theta_\Gamma(r) &=& k_\Gamma r + \frac{Z}{k_\Gamma} \log 2k_\Gamma r - \frac{\ell \pi}{2} + \sigma_{\ell}(k_\Gamma), \label{eq:upwc}
\end{eqnarray}
where $k_\Gamma = \sqrt{2(E-E_\Gamma)}$, $Z$ is the residual charge and $\sigma_\ell(k) = \arg \Gamma(\ell + 1 -iZ/k_\Gamma )$ is the Coulomb phase shift.  The continuum functions $f^{\Gamma E}_{\ell m}$ and $h^{\Gamma E}_{\ell m}$ are orthogonalized to all the molecular orbitals, namely, the \emph{internal} orbitals $\phi_i$ used to build the parent ions $\Phi_\Gamma$, as well as the \emph{external} molecular orbitals $\varphi_\sigma$ featured in the scattering functions $F_{\Gamma',\Gamma E}$:
\begin{equation}
\forall i, \sigma,\,\Gamma,\quad \langle \phi_i | f^{\Gamma E}_{\ell m} \rangle =0 \quad{\rm{and}}\quad \langle \varphi_\sigma | f^{\Gamma E}_{\ell m} \rangle =0.
\label{eq:orthogonalization}
\end{equation}

In earlier publications, the internal and external molecular orbitals were referred to as ``target" and ``scattering" orbitals, respectively~\cite{Rescigno95a,Rescigno95b}. In the case of closed channels, the functions $F_{\Gamma',\Gamma E}$ only comprise the contribution from the external molecular orbitals $\varphi_{\sigma}$.  The states $\Theta_\mu$ in~\eqref{eq:MESA_CCExpansion} relax the orthogonality constraint on the $F_{\Gamma'\Gamma E}$ functions and bring in to the close-coupling expansion additional correlation and polarization terms that cannot be expressed as a parent ion augmented by an internal orbital. 

Here, and for further development, it is convenient to employ a Feshbach notation~\cite{FESHBACH1967} for the partitioning of the Kohn wavefunction. Thus, we write $\psi_{\Gamma E}=P\psi_{\Gamma E}+Q\psi_{\Gamma E}$, where, in the photoionization case, $Q$ is the projector onto the reference space, and $P=1-Q$ projects onto the space spanned by the product of target channel and external orbitals, as well as open target channels and free functions. Since the $P$-space also comprises closed channels, it contributes to the description of Feshbach resonances.

Finally, the coefficients $d_{\mu}^{\Gamma E}$, $c_\sigma^{\Gamma',\Gamma E}$, and the elements of the $T$ matrix in Eqs. (\ref{eq:MESA_CCExpansion}) and (\ref{eq:MESA_ScatteringExpansion}) are treated as variational parameters and obtained by requiring the functional
\begin{equation}
\label{eq:variational_principle}
[T^{\Gamma'\Gamma E}] = T^{\Gamma'\Gamma E} - 2\int\psi_{\Gamma' E}[H_{{\rm eff}}(E)-E]\psi_{\Gamma' E},
\end{equation}
to be stationary (Kohn variational principle). The effective Hamiltonian within the $P$-space, $H_{{\rm eff}}(E) = H_{PP}+V_{{\rm opt}}(E)$, is computed by adding to the energy operator restricted to the $P$-space, $H_{PP}$, the optical potential due to the excitation to the $Q$ space, $V_{{\rm opt}}(E)=H_{PQ}[E-H_{QQ}]^{-1}H_{QP}$~\cite{Rescigno95a,Rescigno95b}. In its present implementation, $H_{PP}$ does not include exchange integrals that explicitly involve continuum functions. This is a consequence of the separable approximation made in the construction of the optical potential, which becomes increasingly more accurate as the Gaussian basis set approaches completeness in the molecular region~\cite{Rescigno95a}.

\subsection{Computation of Transition Dipole Moments} 
The method to compute the one-photon dipole transition matrix elements $\braOket{\psi_{\Gamma E}}{\mathcal{O}}{\Psi_j}$ between a Kohn function, the dipole operator $\mathcal{O}=\hat{\epsilon}\cdot\sum_{i=1}^{N+1}\vec{r}_i$ expressed in the length form, and a bound state expanded in the reference space of internal orbitals, where $\hat{\epsilon}$ is the light polarization vector and $\vec{r}_i$ are the electron vector positions,
has been described in earlier studies \cite{Rescigno93,Douguet13}. The $\braOket {Q\psi_{\Gamma E}}{\mathcal{O}}{\Psi_j}$ and $\braOket{P\psi_{\Gamma E}}{\mathcal{O}}{\Psi_j}$ matrix elements are evaluated separately. Since $\braOket{Q\psi_{\Gamma E}}{\mathcal{O}}{\Psi_j}$ involves bound molecular orbitals only, it is computed with standard many-body techniques already implemented in the MESA code. The $\braOket{P\psi_{\Gamma E}}{\mathcal{O}}{\Psi_j}$ element, which can be expressed in terms of the one-body transition density matrix between the $P\psi_{\Gamma E}$ and the $\Psi_j$ state~\cite{Rescigno95a,McWeenyBook}, is evaluated in two steps. First, the one-particle transition density matrix between the $\Psi_j$ state and channel functions of the form $\hat{\mathcal{A}}[\Phi_\Gamma\varphi_\sigma]$, where $\varphi_\sigma$ is an arbitrary external orbital, is computed,
\begin{eqnarray}
\label{eq:density_matrix}
 \rho^{j\Gamma\sigma}(\mathbf{r}_{1},\mathbf{r}'_{1})&&=\sum_{\zeta_1}\int d\mathbf{x}_{2}\cdots d\mathbf{x}_{N+1}\,\Psi_j(\mathbf{x}_{1},\cdots,\mathbf{x}_{N+1})\times\nonumber\\
&&\hat{\mathcal{A}}\,[\Phi_\Gamma(\mathbf{x}'_{1},\cdots,\mathbf{x}_{N};\zeta_{N+1})\,\varphi_\sigma(\mathbf{r}_{N+1})]_{\zeta_1'=\zeta_1}.
\end{eqnarray}
The only non-vanishing elements of  $\rho^{j\Gamma\sigma}$ (\ref{eq:density_matrix}) are those between an internal orbital, $\phi_i$, and the external orbital $\varphi_\sigma$. Furthermore, these matrix elements do not depend on the latter since $\varphi_\sigma$ is not represented in $\Phi_{\Gamma}$: $\forall \sigma,\sigma'$, \quad $\rho^{j\Gamma}_{i}=\braOket{ \phi_i} { \rho^{j\Gamma\sigma} } {\varphi_\sigma} = \braOket{ \phi_i}{ \rho^{j\Gamma\sigma'} } {\varphi_{\sigma'} }$. In particular, they keep the same value even if $\varphi_\sigma$ is a free function as long as that function has been orthogonalized to the internal orbitals.  Therefore, $\braOket{P\psi_{\Gamma E}}{\mathcal{O}}{\Psi_j}$ can be computed from $\rho^{j\Gamma}_{i}$, $c^{\Gamma',\Gamma E}_\sigma$, and the one-electron transition dipole moments $\braOket{\phi_i}{\mathbf{o}}{\varphi_\sigma}$, $\braOket{\phi_i}{\mathbf{o}}{f^{\Gamma E}_{\ell m}}$, and $\braOket{\phi_i}{\mathbf{o}}{h^{\Gamma E}_{\ell m}}$. The latter elements are evaluated using the same adaptive three-dimensional grid used to compute any other bound-free and free-free integral appearing in the CK method~\cite{McCurdy89}. The current version of the CK method employs the length form of the dipole operator ${\mathbf{o}=\hat{\epsilon}\cdot\vec{r}}$, with $\vec{r}$ the electron vector position. Extension to the velocity form is beyond the scope of the present paper and will be considered for future publications.

\subsubsection{Computation of free-free transition matrix elements}
\label{sec:free-free}
The procedure to compute the transition dipole moments $\mathcal{O}_{\Gamma E,\Gamma' E'}=\braOket{\psi_{\Gamma E}}{\mathcal{O}}{\psi_{\Gamma' E'}}$ between two Kohn functions involves the evaluation of $\braOket{ Q\psi_{\Gamma E}}{\mathcal{O}}{Q \psi_{\Gamma' E'}}$, $\braOket{Q \psi_{\Gamma E}}{\mathcal{O}}{P\psi_{\Gamma' E'}}$, and $\braOket{ P \psi_{\Gamma E}}{\mathcal{O}}{P \psi_{\Gamma' E'}}$. The terms involving the $Q$-space functions can be computed as before, after replacing the real expansion coefficients on the CSFs, $\Theta_\mu$, associated with a correlated bound function $\Psi_j$, by the complex Kohn coefficients $d^{\Gamma E}_\mu$ associated with the Kohn function (\ref{eq:MESA_CCExpansion}). On the other hand, the evaluation of $\braOket{P\psi_{\Gamma E}}{\mathcal{O}}{P\psi_{\Gamma' E'}}$ requires some additional care.

Because the scattering functions $F_{\Gamma',\Gamma E}$ are chosen to be orthogonal to all the molecular orbitals that appear in $\Phi_\Gamma$ , the $P$-space transition dipole matrix elements may be expressed as,
 \begin{eqnarray}
 \label{eq:overlap}
 \braOket{P\psi_{\Gamma E}}{\mathcal{O}}{P\psi_{\Gamma' E'}}&=&\sum_{\Delta\Lambda}
 \braOket{\Phi_{\Delta} F_{\Delta,\Gamma E}}{\mathcal{O}}{\Phi_{\Lambda}F_{\Lambda,\Gamma' E'}}\nonumber\\
 &=&\sum_{\Delta\Lambda}\braOket{ \Phi_{\Delta}}{O|\Phi_{\Lambda}}
\braOket{F_{\Delta,\Gamma E}}{F_{\Lambda,\Gamma' E'}} \nonumber\\
&+&\delta_{\Delta \Lambda} \langle F_{\Delta,\Gamma E} | \mathbf{o} | F_{\Lambda,\Gamma' E'} \rangle,
 \end{eqnarray}
where $\braOket{ \Phi_{\Delta}}{O}{\Phi_{\Lambda}}$ represents transitions between ionic channels $\Lambda$ and $\Delta$, and $\braOket{ F_{\Delta,\Gamma E}}{\mathbf{o}}{F_{\Lambda,\Gamma' E'}}$ represents one-electron transition from a single continuum associated with a common ionic factor.

The overlap and dipole terms in (\ref{eq:overlap}) take the form
\begin{widetext}
  \begin{eqnarray}
 \braket{F_{\Delta,\Gamma E}}{F_{\Lambda,\Gamma' E'}} &=& \sum_\sigma c_\sigma^{\Delta,\Gamma E*}c_\sigma^{\Lambda,\Gamma' E'}
 +\delta_{\Gamma,\Gamma'}\braket{ f^{\Gamma E}_{\ell,m}}{f^{\Gamma' E'}_{\ell',m'}}
 +\sum_{\ell'',m''}T^{\Lambda\Gamma' E'}_{\ell'' m'',\ell' m'}\braket{ f^{\Gamma E}_{\ell,m}}{h^{\Lambda E'}_{\ell'',m''}}\nonumber\\
& +&\sum_{\ell'',m''}T^{\Delta\Gamma E*}_{\ell'' m'',\ell m}\braket{h^{\Delta E}_{\ell'',m''}}{f^{\Gamma' E'}_{\ell',m'}}
 +\sum_{\ell_1,m_1}\sum_{\ell_2,m_2}T^{\Delta\Gamma E*}_{\ell_1 m_1,\ell m}T^{\Lambda\Gamma' E'}_{\ell_2 m_2,\ell' m'}
\braket{h^{\Delta E}_{\ell_1,m_1}}{h^{\Lambda E'}_{\ell_2,m_2}};\label{eq:free-free-overlap}\\
 \braOket{F_{\Delta,\Gamma E}}{\mathbf{o}}{F_{\Lambda,\Gamma' E'}}&=&  \sum_{\sigma,\sigma'} c_\sigma^{\Delta,\Gamma E*}c_{\sigma'}^{\Lambda,\Gamma' E'}\braOket{\varphi_\sigma}{\mathbf{o}}{ \varphi_{\sigma'}}
 +\sum_{\sigma}\Big [c_\sigma^{\Delta,\Gamma E*}\braOket{\varphi_\sigma}{\mathbf{o}}{f^{\Gamma' E'}_{\ell',m'}}+ c_\sigma^{\Lambda,\Gamma' E'} \braOket{ f^{\Gamma E}_{\ell,m}}{\mathbf{o}}{\varphi_\sigma}\Big]\nonumber\\
 &+& \delta_{\Gamma,\Gamma'}\braOket{ f^{\Gamma E}_{\ell,m}}{\mathbf{o}}{f^{\Gamma' E'}_{\ell',m'}}+\sum_\sigma\sum_{\ell'',m''}\Big [T^{\Delta\Gamma E*}_{\ell'' m'',\ell m} c_\sigma^{\Lambda,\Gamma' E'}\braOket{ h^{\Gamma E}_{\ell'',m''}}{\mathbf{o}}{\varphi_\sigma}+ c_\sigma^{\Delta,\Gamma E}T^{\Lambda\Gamma' E'}_{\ell'' m'',\ell' m'}\braOket{\varphi_\sigma}{\mathbf{o}}{h^{\Gamma' E'}_{\ell'',m''}}\nonumber\\
 &+&T^{\Lambda\Gamma' E'}_{\ell'' m'',\ell' m'}\braOket{ f^{\Gamma E}_{\ell,m}}{\mathbf{o}}{h^{\Gamma' E'}_{\ell'',m''}}+T^{\Delta\Gamma E*}_{\ell'' m'',\ell m}\braOket{ h^{\Gamma E}_{\ell'',m''}}{\mathbf{o}}{f^{\Gamma' E'}_{\ell',m'}} \Big]\nonumber\\ &+&\sum_{\ell_1,m_1}\sum_{\ell_2,m_2}T^{\Delta\Gamma E*}_{\ell_1 m_1,\ell m}T^{\Lambda\Gamma' E'}_{\ell_2 m_2,\ell' m'}
\braOket{ h^{\Gamma E}_{\ell_1,m_1}}{\mathbf{o}}{h^{\Gamma' E'}_{\ell_2,m_2}}.\label{eq:free-free-dipole}
 \end{eqnarray}
 \end{widetext}
 Here, we follow the prescription of Rescigno and Orel \cite{PhysRevA.41.1695} and use in Eqs. (\ref{eq:free-free-overlap}) and (\ref{eq:free-free-dipole}) the trial value of the $T$-matrix elements, i.e., the elements appearing on the right hand side of \hbox {Eq. (\ref{eq:variational_principle})}, as dictated by the variational principle~\cite{PhysRevA.41.1695}

The overlap between orthogonalized asymptotic functions in (\ref{eq:free-free-overlap}) and (\ref{eq:free-free-dipole}) are decomposed into two parts. For the case of two outgoing functions, the overlap is written as 
\begin{equation}
 \label{eq:OverlapSurfaceTerm1}
\braket{ h^{\Gamma E}_{\ell,m}}{h^{\Gamma' E'}_{\ell',m'}}=\braket{h^{\Gamma E}_{\ell,m}}{h^{\Gamma' E'}_{\ell',m'}} _{[0,R_b]}+\braket{ h^{\Gamma E}_{\ell,m}}{h^{\Gamma' E'}_{\ell',m'}}_{[R_b,\infty]}.
\end{equation}
The first part, $\braket{h^{\Gamma E}_{\ell,m}}{h^{\Gamma' E'}_{\ell',m'}}_{[0,R_b]}$, is the overlap evaluated up to a fixed boundary radius $R_b$ and computed with the three-dimensional grid of McCurdy and Rescigno \cite{McCurdy89}. 
Then, using the identity:
\begin{eqnarray}
 &(\epsilon-\epsilon') \braket{\varphi}{\varphi'}_{[R_b,\infty]} = \braket{ H \varphi}{\varphi'}_{[R_b,\infty]}-\braket{ \varphi}{H\varphi'}_{[R_b,\infty]}\nonumber\\
 &= \frac{1}{2}\braket{\varphi}{r^{-1}\partial_r^2 r\varphi'}_{[R_b,\infty]}
 -  \frac{1}{2}\braket{ r^{-1}\partial_r^2 r \varphi}{\varphi'}_{[R_b,\infty]}, 
 \end{eqnarray}
 it can be shown that the second part of the integral can be computed using the following expression
 \begin{equation}
 \label{eq:OverlapSurfaceTerm2}
 \braket{ h^{\Gamma E}_{\ell,m}}{h^{\Gamma' E'}_{\ell',m'}}_{[R_b,\infty]} = \frac{1}{2}\frac{\mathcal{P}\mathcal{S}(R_b)}{\epsilon_\Gamma-\epsilon'_{\Gamma'}}+\frac{\pi}{4}\delta_{\Gamma',\Gamma}\delta(\epsilon_\Gamma-\epsilon'_{\Gamma'}),
 \end{equation}
 where $\mathcal{P}$ denotes the principal value. The prefactor $\pi/4$ in front of the delta function arises from the proper normalization of the asymptotic outgoing function in (\ref{eq:outgoing}), and the surface term ${\cal S}(R)$, evaluated at $R$, is given by
  \begin{equation}
 {\mathcal S}(R)=R[h^{\Gamma E*}_{\ell,m}\partial_rrh^{\Gamma' E'}_{\ell',m'}-h^{\Gamma' E'*}_{\ell',m'}\partial_r rh^{\Gamma E}_{\ell,m}]_R.
 \end{equation}
 We have verified numerically that the overlaps in (\ref{eq:OverlapSurfaceTerm1}) are independent of the choice of $R_b$ to a high level of precision.

We now turn towards the computation of one-electron free-free transition dipole moments, e.g., terms such as  $\braOket{ f^{\Gamma E}_{\ell,m}}{\mathbf{o}}{h^{\Gamma' E'}_{\ell',m'}}$. In the velocity form, the dipole matrix elements can be decomposed as in (\ref{eq:OverlapSurfaceTerm1}), with the advantage that the boundary term can be expressed in analytical form as $r \rightarrow \infty$ \cite{PhysRevA.90.012506}. The transition dipole moment in the length form, $\braket{f^{\Gamma E}_{\ell,m}}{\hat{\epsilon}\cdot\vec{r}{|h^{\Gamma' E'}_{\ell',m'}}}$, on the other hand, requires additional caution~\cite{Madajczyk89,Veniard90,Komninos12}, as it exhibits a second order pole at $E=E'$.  In hydrogen, the singular part of the dipole matrix elements is known analytically \cite{Madajczyk89,Piraux90}.  Komninos~\textit{et al.}~\cite{Komninos12} have reported that the regularized dipole operator $D_{SF}(r)=r$, for $r< r_0$, and $D_{SF}(r)=r_0$, for $r\ge r_0$, represents a good approximation as long as $r_0=3\lambda/8$, where $\lambda$ is the laser wavelength. This finding indicates that a smooth regularization of the dipole kernel can be a practical way of circumventing the \emph{on shell} singularity.

\begin{figure}[t]
\includegraphics[width=8.5cm]{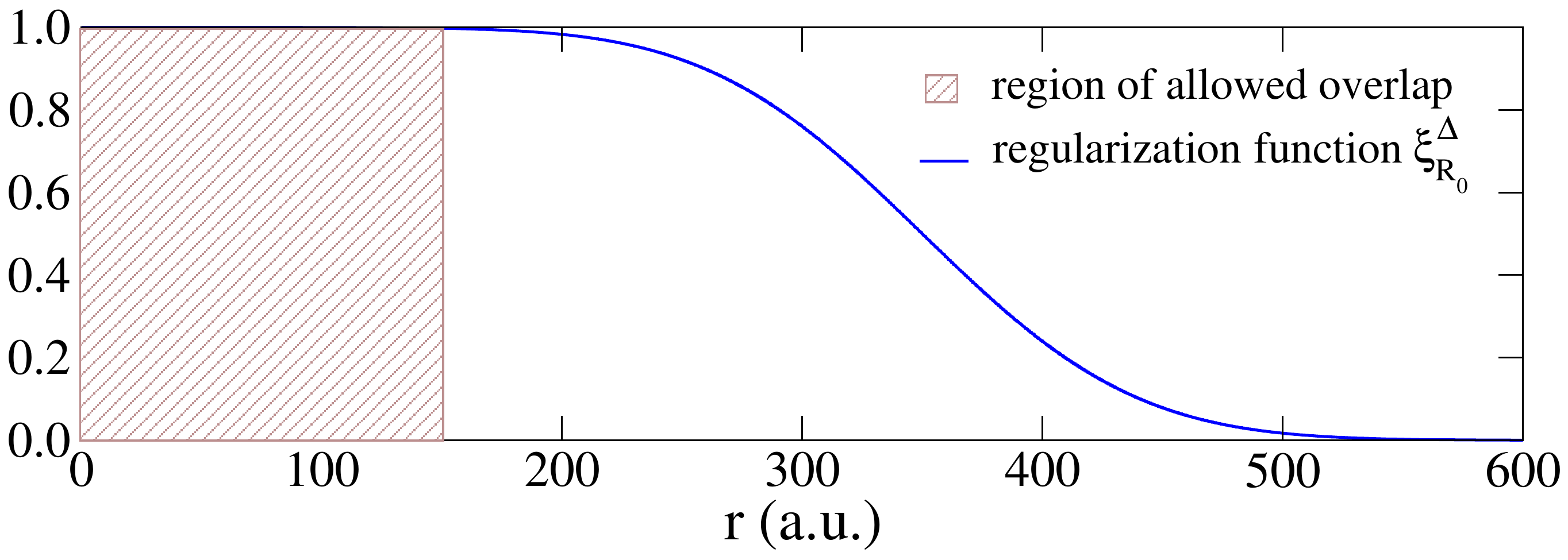}\\
\caption{Function $\xi^\Delta_{R_0}(r)$ with $R_0=350$~a.u. and  $\Delta=100$~a.u. employed in this study to regularize the free-free dipole moments expressed in the length form. 
The dashed domain represents the region where bound and free functions can overlap. }
\label{fig:1}
\end{figure}

In the present case, we have chosen to regularize the dipole matrix elements as 
\begin{equation}
\label{eq:dipole_integral}
\braOket{f^{\Gamma E}_{\ell,m}}{\hat{\epsilon}\cdot\vec{r}}{h^{\Gamma' E'}_{\ell',m'}}=\hat{\epsilon}\cdot\int f^{\Gamma E}_{\ell,m}(\vec{r})~\vec{r}~h^{\Gamma' E'}_{\ell',m'}(\vec{r})\xi^\Delta_{R_0}(r)d^3r,
 \end{equation}
 where $\xi^\Delta_{R_0}(r)$, exemplified in Fig. \ref{fig:1}, is a smooth step function that transitions from $\approx 1$ to $\approx 0$ in an interval  centered at $r=R_0$ and with characteristic width $\Delta$, 
 \begin{equation}
 \label{eq:regularization_function}
\xi^\Delta_{R_0}(r)=\frac{1}{2}-\frac{1}{2}{\rm erf}\left(\frac{r-R_0}{\Delta}\right),
\end{equation} 
where $\mathrm{erf}(x)$ is the error function, \hbox{${\rm erf}(x)=\frac{2}{\sqrt \pi}\int_{0}^x e^{-t^2}dt$}~\cite[Eq~7.2.1]{NIST:DLMF}.
The parameters  $R_0$ and $\Delta$ in (\ref{eq:regularization_function}) should be set such that $\xi^\Delta_{R_0}(r)\sim1$ to high accuracy in the region where Gaussian functions assume non-negligible values. This condition ensures consistency with the orthogonalization procedure between free and bound functions. The parameter $R_0$ provides a measure of the size of the region where net photon exchange occurs. Large values of $r$ are not involved with effective photon exchange, but solely with the oscillatory motion of the electron driven by the external field. As $R_0$ is increased, the regularized free-free transition dipole moment gradually approaches the $(\epsilon_\Gamma-\epsilon'_{\Gamma'})^{-2}$ divergence. Any finite-resolution observable computed using this regularization procedure is expected to converge to a physically sound value in the limit of $R_0\to\infty$ (or, optionally, $\Delta\to\infty$ and $\Delta/R_0\to 0$). The role of the width $\Delta$ is to ensure a smooth decay and hence to accelerate spectral convergence, since the artifacts associated to the regularization exhibit a spectral width of the order of $1/\Delta$. In practice, the validity of this regularization procedure is assessed through the convergence of the final results with respect to $R_0$ and $\Delta$.

\subsection{Two-photon dipole transition matrix elements}

The two-photon dipole transition matrix elements (2PTME), from the ground state of an atom or molecule to a final state in the continuum associated with a channel $\Gamma$, have the following expression
\begin{equation}
{\cal M}_{\Gamma E,g}(\omega)=\langle \psi_{\Gamma E}|\mathcal{O}G^+(\omega_g+\omega)\mathcal{O}|\Psi_g\rangle,
\end{equation}
where $G^+(\omega)=(\omega-H+i0^+)^{-1}$ is the retarded resolvent of the field-free hamiltonian $H$, $|\Psi_g\rangle$ is the ground state wavefunction of the system with energy $E_g=\hbar\omega_g$, and $\mathcal{O}$ is the dipole operator in the length form. One way to evaluate 2PTME is to expand the intermediate resolvent in a complete set of states. Assuming that such an expansion can be limited to bound and single-ionization scattering states, the 2PTME is expressed in the following form
\begin{equation}
\label{eq:2PTME}
{\cal M}_{\Gamma E,g}(\omega)=\sum_{\Gamma'}\SumInt dE'\frac{\mathcal O_{\Gamma E,\Gamma' E'}\mathcal O_{\Gamma' E',g}}{\omega+\omega_g-E'+i0^+},
\end{equation}
where the integral includes a summation over bound state energies as well as an integration over the continuum levels. 
The index $\Gamma$ is used to characterize electronic bound states below the first ionization threshold as well as the degenerate continuum levels.
The evaluation of ${\cal M}_{\Gamma E,g}(\omega)$ in Eq. (\ref{eq:2PTME}) using variational scattering functions is the principal aim of the present study. From a numerical standpoint, the elements ${\mathcal O}_{\Gamma' E',g}$ and ${\mathcal O}_{\Gamma E,\Gamma' E'}$ must be computed using an energy mesh adequate to perform a quadrature in the continuum.  It is important to pay particular attention close to resonances, where the energy denominators become small and there are large contributions to the sum over states. In order to ensure the accuracy of  (\ref{eq:2PTME}), the dipole moments are finally interpolated on a finer energy grid and 
the {\textsc{quadpack}} package~\cite{quadpack} is employed to evaluate the principal value $\mathcal P$ of the integral.

The ionization amplitudes in the perturbative regime can be readily evaluated from ${\cal M}_{\Gamma E,g}(\omega)$. In the present situation for which we consider a linearly polarized electric field to fix ideas, $E(t)={\bm E(t)}\cdot\hat z$, with Fourier transform (FT) $\tilde F(\omega)=(2\pi)^{-1}\int E(t)\exp(i\omega t)dt$, the two-photon 
amplitude takes the well known form~\cite{PhysRevA.93.023429}
\begin{equation}
\label{eq:tw-photon-amplitude}
\mathcal A^{(2)}_{\Gamma E,g}=-i\int \tilde F(\omega_{E,g}-\omega)\tilde F(\omega){\cal M}_{\Gamma E,g}(\omega)d\omega,
\end{equation}
where $\omega_{E,g}=E-E_g$. It is straightforward to obtain results for various characteristics of the pulses (e.g. harmonic wavelength, light intensity, or pump-probe time-delay) as one only needs to recompute the FT of the field and evaluate the simple integral (\ref{eq:tw-photon-amplitude}). 

\section{Results and discussion}
\label{sec:2}

To benchmark the present extension of the CK method to compute continuum-continuum transitions in poly-electronic systems, we consider the two-photon photoionization of the helium atom in the proximity of the  $2s2p$~{$^1$P$^o$} autoionizing state. This process is sufficient to illustrate the main aspects of the approach, and has already been the subject of extensive theoretical and experimental investigations in the past. We will consider only small-size, but quite accurate calculations to test the new approach. It should be noted that the accuracy of the results can in principle be systematically improved by extending the size of the close-coupling expansion in Eq. (\ref{eq:MESA_CCExpansion}).

\subsection{Description of the scattering calculations}

For the short-range orbitals, we employ the \texttt{cc-pVQZ} basis set of Woon and Dunning \cite{Dunning94}, complemented it with $s$ and $p$ diffuse orbitals with exponents 0.05 and 0.02. The basis set is then used to compute He$^+$ orbitals by diagonalizing the one-electron Hamiltonian. A well-known difficulty in photoionization is the accurate description of both the $N$ and $(N+1)$-electron system from a single set of atomic or molecular orbitals. Here, we improve the description of the $1s^2$ ground state by including the $1s$ Hartree-Fock orbital in the construction of the orbital space that is used for both the ions and the neutral.
In the calculation of the scattering states, besides the $1s$ He$^+$ parent ion, we include also the $2s$ and $2p$ excited channels, whose energies are virtually exact. The two-electron reference space is built out of four $s$ orbitals and two $p$ orbitals, leading to a ground state energy for the neutral atom of $-2.88040$ a.u., which is very close to the Hartree-Fock limit and not far from the exact non-relativistic value ($-2.90372$ a.u.).
 The resulting close-coupling expansion in (\ref{eq:MESA_CCExpansion}) includes a total of 145 CSFs. 
 
 As most quantum-chemistry codes, \MESA~is able to account for molecular symmetry in terms of the $D_{2h}$ point group and its subgroups. In the case of linearly polarized light along the $\hat{z}$ axis, therefore, the He ground state belongs to the $A_g$ irreducible representation, and for the excited states we need only consider the states with $A_g$ and $B_{1u}$ symmetry. To simplify the discussion, in the following we will hide this inessential technicality and refer to state symmetry in their proper $SO(3)$ representation $S$, $P$, and $D$. 
\begin{figure}[htbp]
\includegraphics[width=8.0cm]{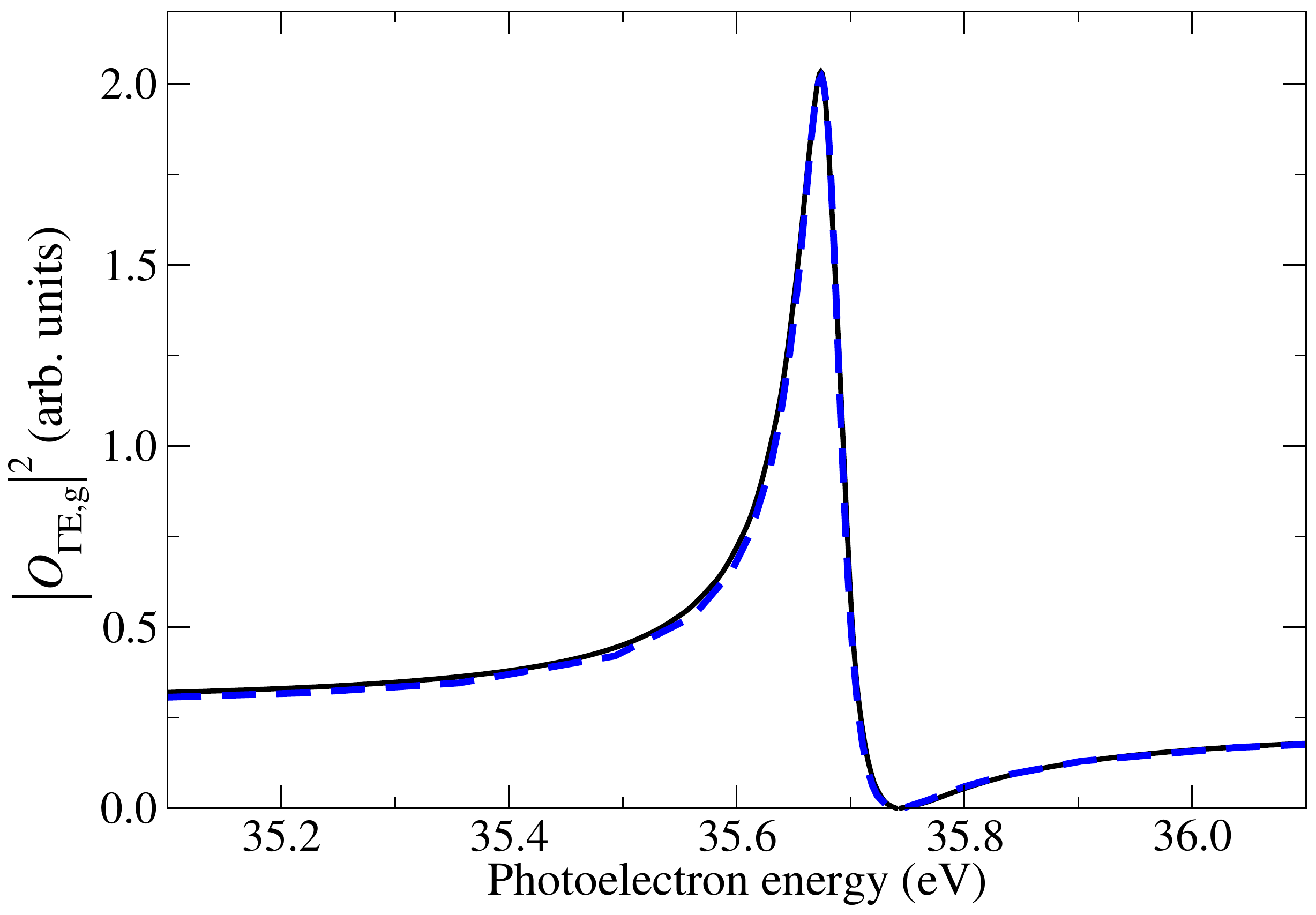}\\
\caption{Square of the bound-free transition dipole moment  $|O_{\Gamma E,g}|^2$ between He ground state $\Psi_g$ and continuum state $\Psi_{\Gamma E}$,
calculated using the Kohn method (black solid line) and  \STOCK~(blue dashed line).}
\label{fig:2}
\end{figure}

The calculations used for comparison are based on the atomic {\STOCK}~code, which, for single-photon ionization processes in stationary conditions, has proven to provide reliable results~\cite{STOCK2013,Chew18,Marante17}. The present calculations are based on a time-dependent extension of the \STOCK~code \cite{Argenti18NEWSTOCK}. For the present purpose, it suffices to say that \STOCK~has been used to construct a close-coupling basis equivalent to the one employed in the CK calculation, but in which the radial part of both the localized and the continuum orbitals is expressed in terms of a B-spline basis~\cite{Bachau2001} with asymptotic node spacing $\Delta r=0.4$~a.u., reaching a maximum radius of approximately $600$~a.u. The parent-ion orbitals are virtually exact. The scattering states $\psi_{\alpha E}^-$, where $\alpha$ is any channel open at the total energy $E$, are obtained by solving the Lippmann-Schwinger equation, with incoming boundary conditions at the edge of the quantization box, using a method equivalent to the one explained in Sec.3.4 of~\cite{MaranteJCTC2017}. Thanks to the small number of channels, the solution of the scattering problem is, within the chosen truncated close-coupling space, also virtually exact. For the purpose of comparing with the CK calculation, the continuum-continuum matrix elements in \STOCK~are estimated using the same regularization procedure discussed in Sec.~\ref{sec:free-free}. To compute the effect of a sequence of pulses on the ground state of the atom, the TDSE is numerically integrated in the spectral space of the field-free Hamiltonian $H_0$ projected on the B-spline close-coupling basis, starting from the {$^1$S$^e$} ground state and using the second-order unitary exponential propagator 
\begin{equation}
U(t+dt,t)=e^{-iH_0 dt/2}e^{-iH_I(t+dt/2)dt}e^{-iH_0 dt/2}e^{-iV_{CAP} dt},
\end{equation}
where $H_I(t)=\alpha \vec{A}(t)\cdot\vec{p}$ is the dipole interaction operator in velocity form and $V_{CAP}$ is a complex-absorbing potential that prevents unphysical reflection of the photoelectron wavepacket from the box boundary. Since the goal of the present calculation is to compare the result of two separate methods, we have restricted the configuration space to only the {$^1$S$^e$} and {$^1$P$^o$} symmetries, leaving out the {$^1$D$^e$} symmetry. The time-dependent wavefunction $\Psi(t)$ is expressed in the spectral basis of $H_0$. As a consequence, the action of $H_0$ on the propagating wavefunction is exact, whereas the action of the dipolar Hamiltonian is calculated with an iterative Krylov-space method. Once the pulse has terminated, the asymptotic photoelectron distribution is obtained by projecting the two-electron wavepacket on scattering solutions fulfilling incoming boundary conditions.
\begin{figure*}[htbp]
\begin{center}
$ \begin{array}{cc}
\includegraphics[width=8.cm]{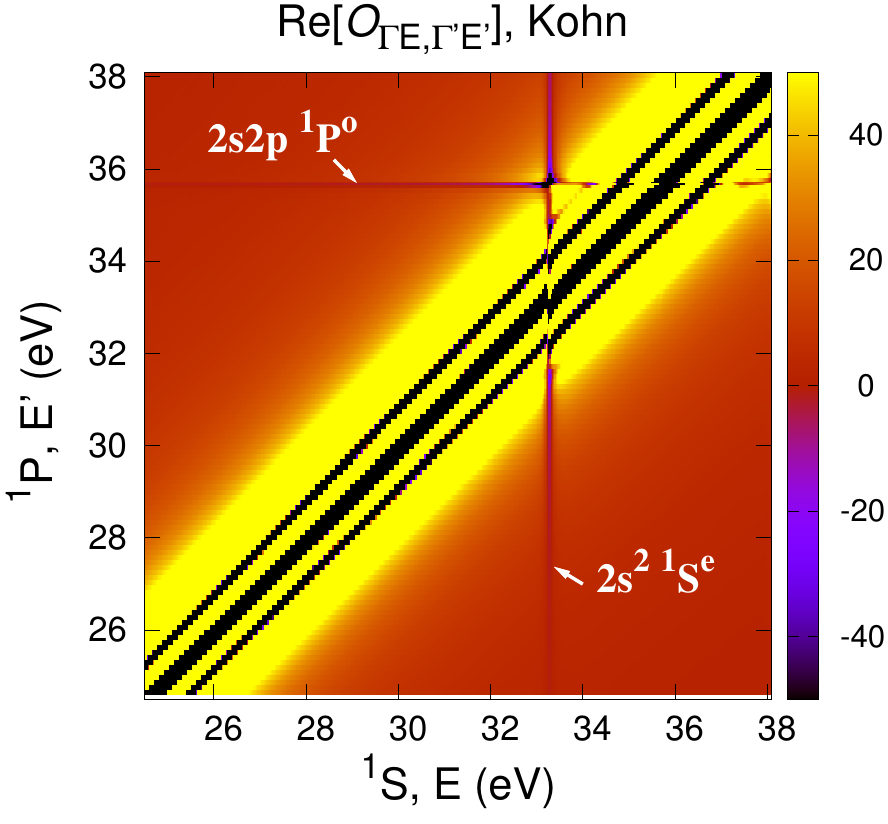}~\includegraphics[width=8.cm]{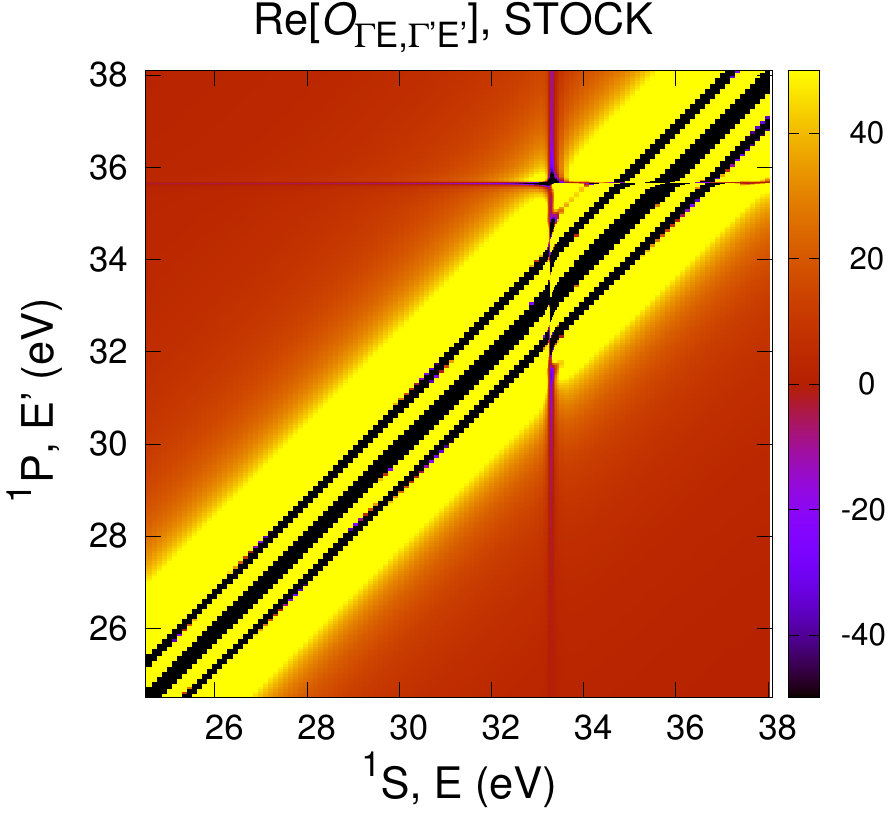}\\
\includegraphics[width=8.cm]{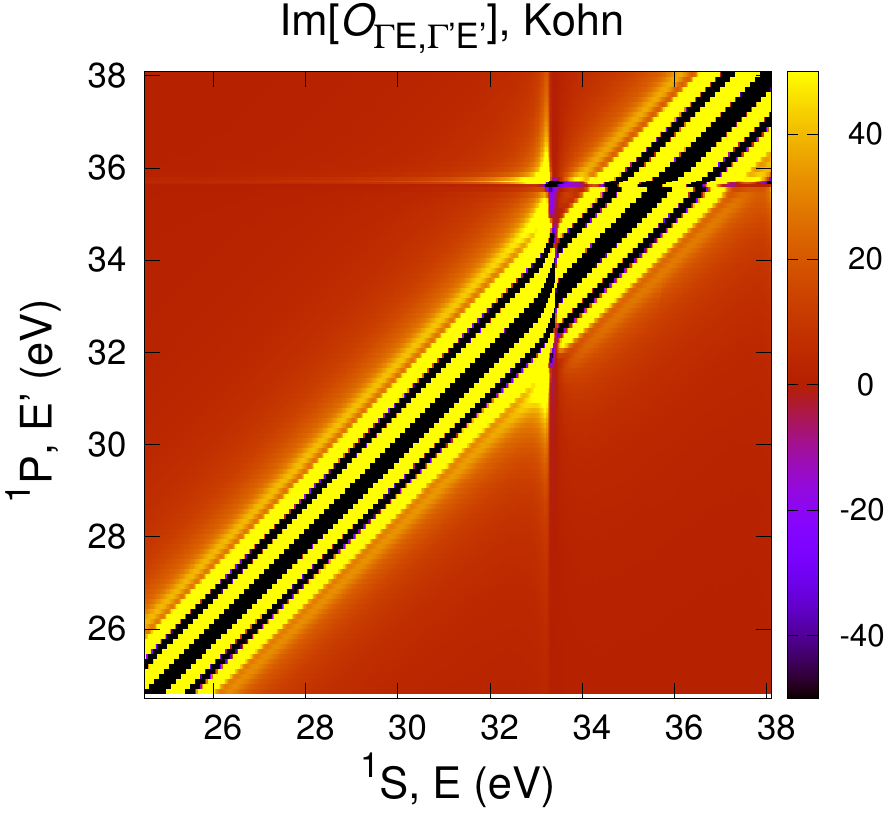}~\includegraphics[width=8.cm]{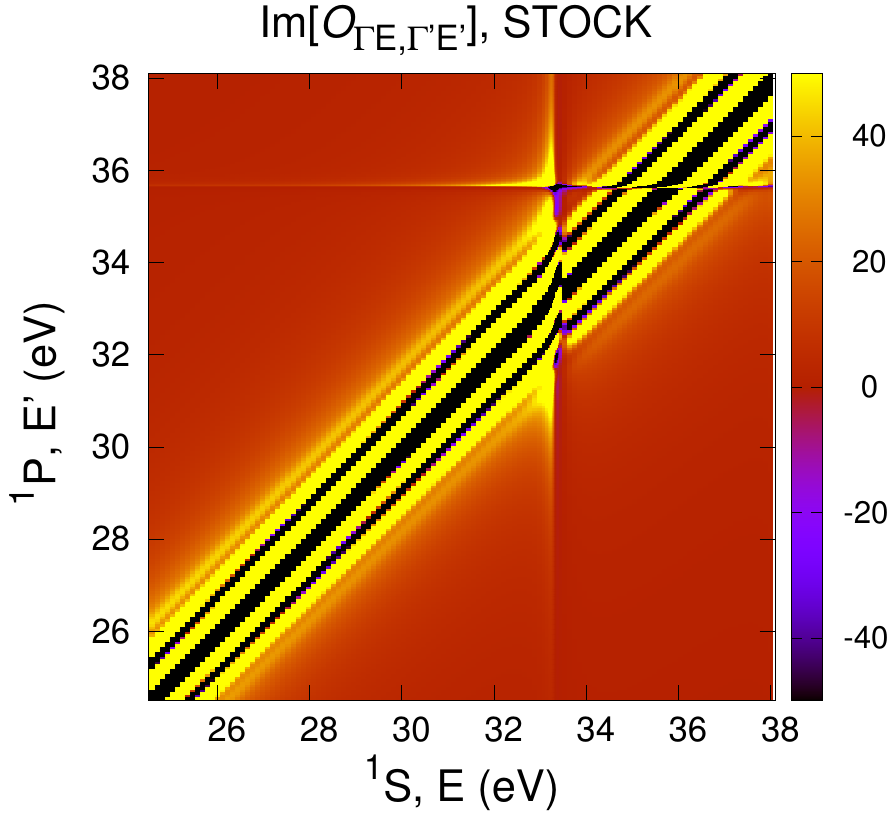}
\end{array}$
\caption{Real (upper panels) and imaginary (lower panels) parts of the free-free transition dipole moments $\mathcal O_{\Gamma E, \Gamma' E'}$ as a function of $E'$ and $E$, from the $^1P$ to $^1S$ symmetries, calculated in the length gauge, using the Kohn (left panels) and the velocity gauge \STOCK~(right panels) methods (see text for details).}
\label{fig:3}
\end{center}
\end{figure*}

\subsection{Bound-free, free-free and two-photon dipole transition elements}
\subsubsection{Bound-free transition dipole moments}
The consistency between single-channel continua, at energies below the $2s/2p$ threshold, computed with the MESA+CK and with the \STOCK~code can be confirmed by comparing the corresponding dipole transition matrix elements. Figure~\ref{fig:2} shows the bound-free transition dipole moment between He ground state and the $^1P$ continuum near the ($2s2p$)$^1P^o_1$ resonance. The resonance position differ by $\sim$13~meV between the two methods, which is a small discrepancy compatible with the difference between the two radial bases (Gaussian functions vs B-splines, in \MESA~and \STOCK, respectively).
The photoionization cross sections computed with the two approaches exhibit excellent agreement once the \MESA~calculation are scaled by an overall factor of~$1.2$, arguably due to the difference with which the transition of the continuum wave function, from short to long radii, is accomplished in the two methods, when a limited number of Gaussian functions in MESA+CK is employed. The cross section profile reproduces well the experimental data~\cite{PhysRevA.29.1901}.

\subsubsection{Free-free transition dipole moments}
We now consider the free-free transition dipole moments $\mathcal O_{\Gamma E,\Gamma' E'}$ obtained in the length form of the dipole operator using the continuum states from either the CK or the \STOCK~code.  The real and imaginary parts of the elements $\mathcal O_{\Gamma E,\Gamma' E'}$ are shown in \hbox{Fig. \ref{fig:3}} for transitions between two continuum states $\Psi_{\Gamma' E'}$ and $\Psi_{\Gamma E}$, with $^1P$ and $^1S$ symmetry, respectively.
We observe a striking agreement between the two completely independent set of calculations. The sharp horizontal and vertical features near 35.6 and 33.2~eV photoelectron energy, correspond to the ($2s2p$)$^1P^o_1$ and ($2s^2$)$^1S^e_1$ resonances, respectively. The dipole moment diverges as $(E-E')^{-2}$ for $E\approx E'$ \cite{Madajczyk89,Piraux90} and the superimposed oscillations near $E\approx E'$ are due to our regularization procedure in Eq. (\ref{eq:dipole_integral}), where we set $R_0=350$~a.u. and $\Delta=100$~a.u (see Fig. \ref{fig:1}). 
This procedure restricts photon exchange to a finite radial region, $r\leq R_0$, which is a satisfactory approximation if the change in radial momentum associated to the absorption or emission of a probe photon $\omega$ is sufficiently large, $(k-k') \gg 1/R_0$, or, alternatively, $\omega \gg \sqrt{E}/R_0$. In the test calculations discussed below, where we combine XUV with either optical or $800$~nm near infrared light, and with the present value of $R_0$, this condition is met and indeed we find good agreement with benchmark calculations. To describe the effect of infrared light with longer wavelength, on the other hand, may require extending $R_0$ to larger values.

The study of He photoionization above the first excited threshold is beyond the scope of this work, however, it is well within the capability of the present approach. 
Highly excited Rydberg states, which are well described in \STOCK, can be included in the Kohn method as well, by adding very diffuse gaussian functions. Indeed, structure calculations of Rydberg states up to principal quantum number $n=5$ were successfully performed using \MESA~\cite{Fonseca14}. 

\subsubsection{Two-photon dipole transition elements}
\label{sec:2b}
The bound-free ${\mathcal O}_{\Gamma E,g}$ and free-free ${\mathcal O}_{\Gamma E,\Gamma' E'}$ transition dipole moments can be combined to compute 
the 2PTME ${\cal M}_{\Gamma E,g}(\omega)$. First, we recast Eq. (\ref{eq:2PTME}) in the following way

\begin{eqnarray}
\label{eq:2PTME-2}
{\cal M}_{\Gamma E,g}(\omega)&=&\sum_{\Gamma'}\mathcal P\int dE'\frac{\mathcal O_{\Gamma E,\Gamma' E'}\mathcal O_{\Gamma' E',g}}{\omega+\omega_g-E'}\nonumber\\
&-&i\pi\sum_{\Gamma'} \mathcal O_{\Gamma E,\Gamma' \omega+\omega_g}\mathcal O_{\Gamma' \omega+\omega_g,g},
\end{eqnarray}
where in our case, the final channel $\Gamma$ corresponds to electronic states with either $^1S$ or $^1D$ symmetry. 
For photoionization leading to He$^+(1s)$ ground state, the final photoeletron angular momentum is thus either $\ell=0$ or $\ell=2$. In the above expression, we discarded the contribution from excited bound states, since it is negligible at the photoelectron energy considered. Computing the principal part of the integral in Eq. (\ref{eq:2PTME-2}) using the {\textsc{quadpack}} \cite{quadpack} package, we have taken an integration interval in $E'$ extending from 26 to 43~eV. We have verified that this energy range is sufficient to ensure the convergence of ${\cal M}_{\Gamma E,g}(\omega)$ in the energy region and light frequency of interest. 

\begin{figure}[t]
\includegraphics[width=8.0cm]{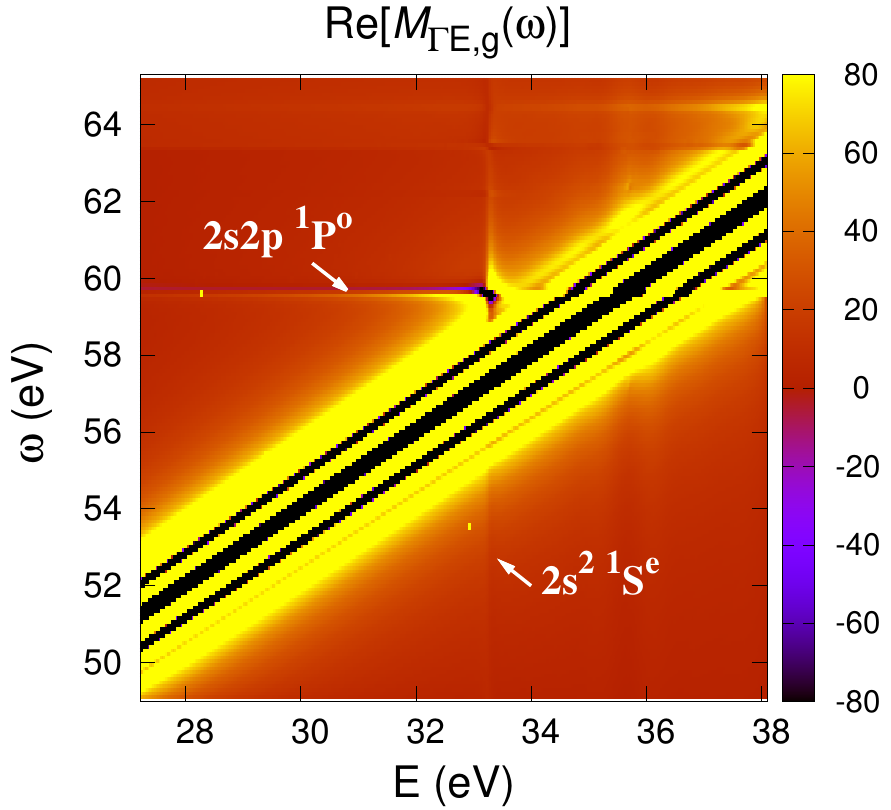}
\caption{Real part of ${\cal M}_{\Gamma E,g}(\omega)$ for a two-photon transition to the final $^1S$ electronic state leading to $^2$He$^+(1s)$ product as a function of the final electron energy $E$ and the photon frequency $\omega$.}
\label{fig:4}
\end{figure}

The real part of ${\cal M}_{\Gamma E,g}(\omega)$ is shown in Fig. \ref{fig:4}. The effect of the  ($2s2p$)$^1P^o_1$ resonance is clearly seen near $\omega=58.6$~eV. One can also observe in the 2PTME the signature of high-lying Rydberg states with $^1P^o_1$ symmetry. The band of singular values of ${\cal M}_{\Gamma E,g}(\omega)$ at $\omega=E-\omega_g$ is associated to the near singularity of $\mathcal O_{\Gamma E,\Gamma' E'}$ for $E\approx E'$. We also recognize the characteristic oscillations due to the regularization procedure (\ref{eq:dipole_integral}).

\subsection{Ionization Schemes and two-photon ionization probability}
\label{sec:2c}
Attosecond spectroscopy enables us to study photoemission in the time domain and access information on ultrafast processes. In the RABBITT technique~\cite{Paul1689}, an XUV attosecond pulse train (APT), generated by the interaction between an active medium and an intense VIS or IR pulse, is combined with a delayed weak replica of the latter and used to ionize a target atom or molecule. In traditional RABBITT, the APT comprises only odd-order harmonics of the fundamental driving laser frequency. In the weak field regime, two distinct ionization pathways interfere at a middle sideband SB$_{2n}$; the first pathway is characterized by the absorption of a photon from the harmonic H$_{2n+1}$ followed by the emission of an IR photon, while the second pathway involves the absorption of a photon from H$_{2n-1}$ followed by the absorption of an IR photon. The amplitudes for these two processes add with different relative phases as a function of the time-delay $\tau$ between the APT and the IR pulse. As a result, the interference between the two amplitudes gives rise to an harmonic beating of the sideband intensity at twice the IR frequency. This technique has been extensively applied to measure photoemission delays in atoms (see \cite{Gruson734} and references therein) and, most recently, to study the photoemission delay anisotropy near a Fano resonance in argon~\cite{Cirelli18}. 
\begin{figure}[t]
\begin{center}
$ \begin{array}{cc}
\includegraphics[width=4.25 cm]{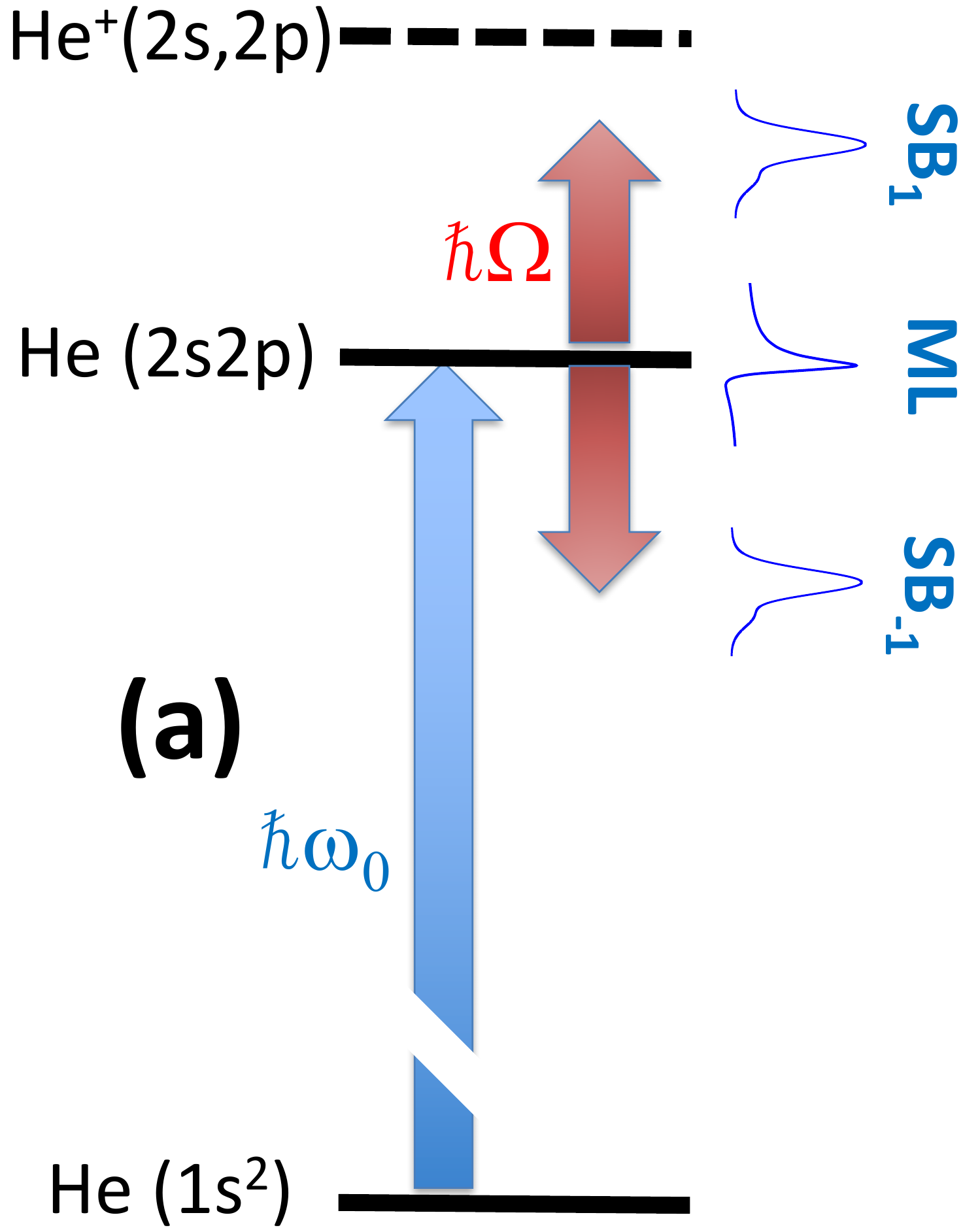}&\includegraphics[width=4.25 cm]{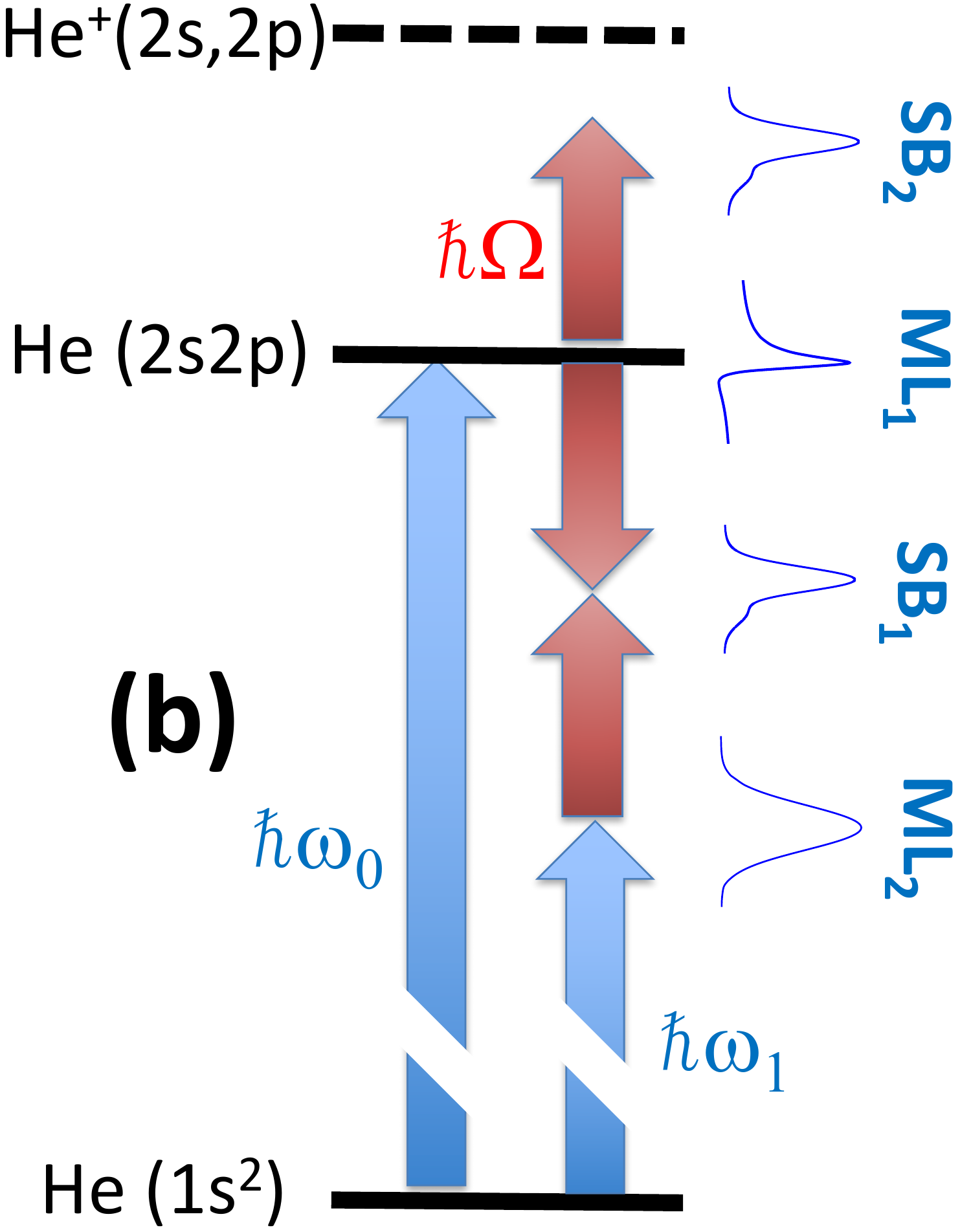}
\end{array}$
\caption{Ionization schemes considered in this study: (a) two-photon ionization with a combined XUV and IR/optical field, and (b)
RABBITT-like scheme with two XUV pulses with frequencies, $\omega_0$ and $\omega$ separated by twice the IR frequency.}\label{fig:5}
\end{center}
\end{figure}
Here, we propose to test the validity of our methodology by considering a similar pump-probe scheme. 

First, we benchmark the accuracy of two-photon ionization amplitudes, which are central quantities in RABBIT spectroscopy, in the presence of just one XUV harmonic, as illustrated in~\hbox{Fig. \ref{fig:5}(a)}. The time-dependent vector potential 
$A(t)=A_{\omega_0}(t) + A_{\Omega}(t)$, with $A(t)=\bm A(t)\cdot \hat{z}$, is formed out of an XUV pulse with central  frequency $\omega_0$, tuned at the ($2s2p$)$^1P^o_1$ resonance,
and an overlapping IR/optical field of frequency $\Omega$ that induces absorption or emission of an additional photon. The  vector potential of the XUV pulse has a Gaussian envelope with $\sigma=5.5$~fs,
\begin{equation}
A_{\omega_0}(t)=A_0\cos(\omega_0 t)\exp\left(\frac{-4t^2\ln2 }{\sigma^2}\right),
\end{equation}
whereas the vector potential of the IR/optical field has a cosine-square envelope, it contains $N$-cycles of the central carrier frequency, and is delayed by $\tau$ with respect to the center of the XUV pulse, 
\begin{equation}
A_{\Omega}(t,\tau)=A_\Omega\cos\left[\Omega (t-\tau)\right]\cos^2\left(\frac{\Omega (t-\tau)}{2N}\right).
\end{equation}
The ionization by the XUV field gives rise to a main line (ML) in the photoelectron spectrum, with a characteristic Fano resonant modulation. The sidebands, for either IR photon emission (SB$_{-1}$) or absorption (SB$_{1}$), shaped by both the two-photon matrix element and the field spectrum, reproduces the resonant profile with smoother features.
\begin{figure}[t]
\begin{center}
$ \begin{array}{c}
\includegraphics[width=8.25 cm]{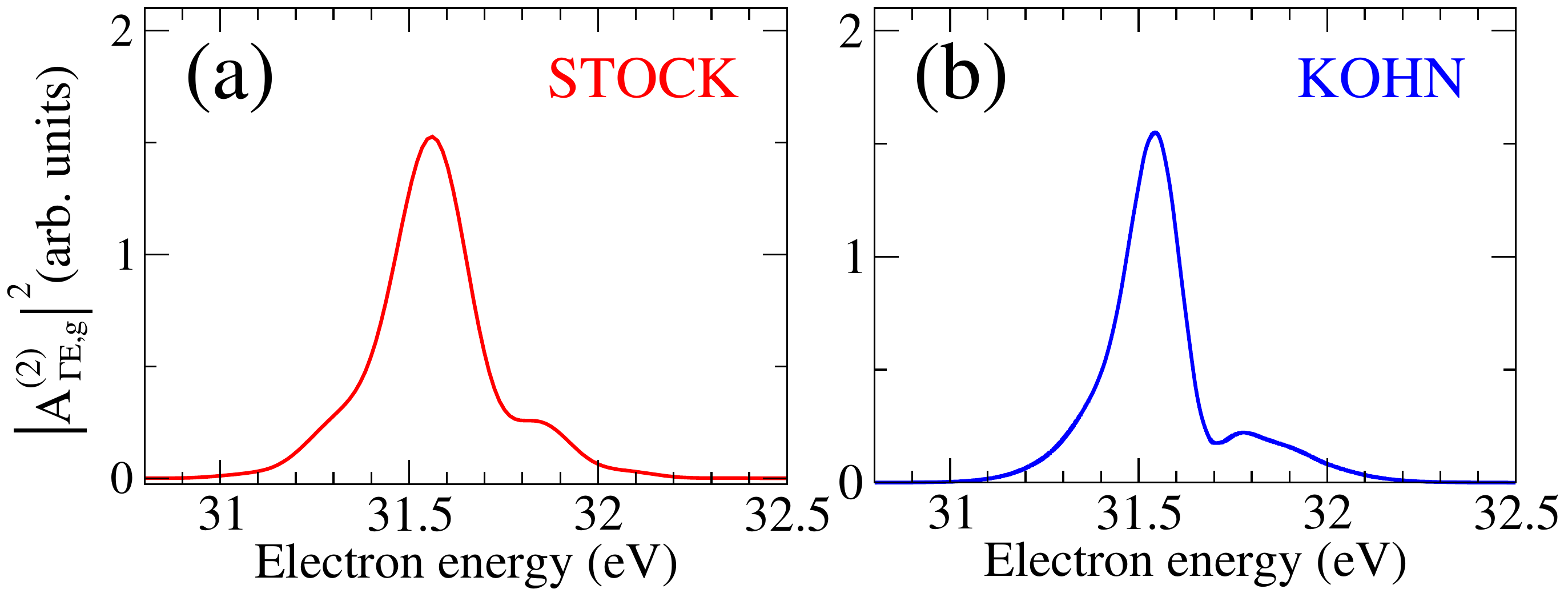}\vspace{-0.0 cm}\\
\includegraphics[width=8.25 cm]{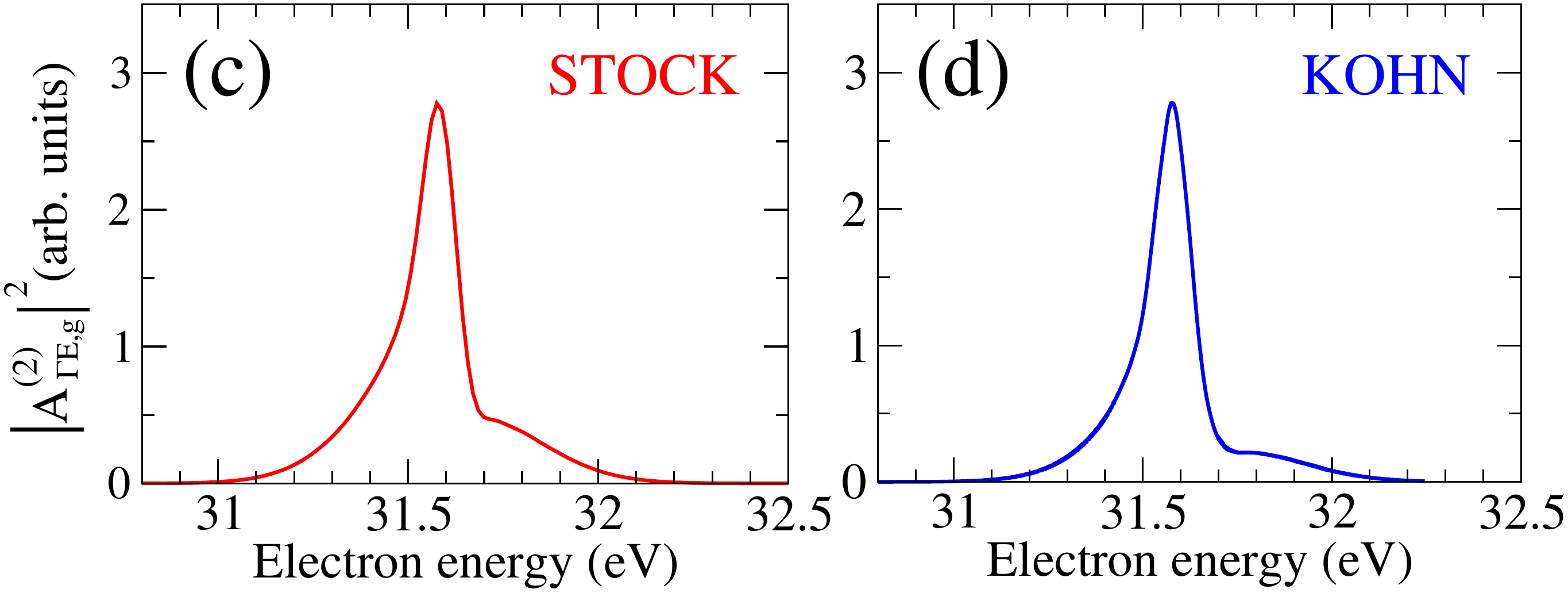}
\end{array}$
\caption{Two-photon ionization probability from helium ground state to the final $\Gamma={}^1S$ state at SB$_{-1}$ calculated
using both \STOCK~and Kohn methods. The optical pulse has intensity $2.5\times10^{11}$~W/cm$^2$, wavelength $300$~nm,
and FWHM $30$~fs in (a) and (b), and $50$~fs in (c) and (d).}\label{fig:6}
\end{center}
\end{figure}

To assess the accuracy of $\mathcal A^{(2)}_{\Gamma E,g}$ calculated in the CK method, we have performed calculations, varying the probe frequencies, their duration, and relative delay. For all calculations, the XUV pulse has a peak intensity of 10$^{11}$W/cm$^2$. Here, we present the ionization probability to the $\Gamma={^1\mathrm{S}}$ state only, i.e., $\ell = 0$, and we rescale the ionization probability to obtain equal maximum of the peaks computed with the \STOCK~and the 2PCK methods. In contrast with the 2PCK method, the TDSE calculations in \STOCK~were performed in  velocity form. Agreement between calculations obtained in different gauges is strong evidence of the robustness of the calculation.

\begin{figure}[b]
\begin{center}
$ \begin{array}{c}
\includegraphics[width=8.25 cm]{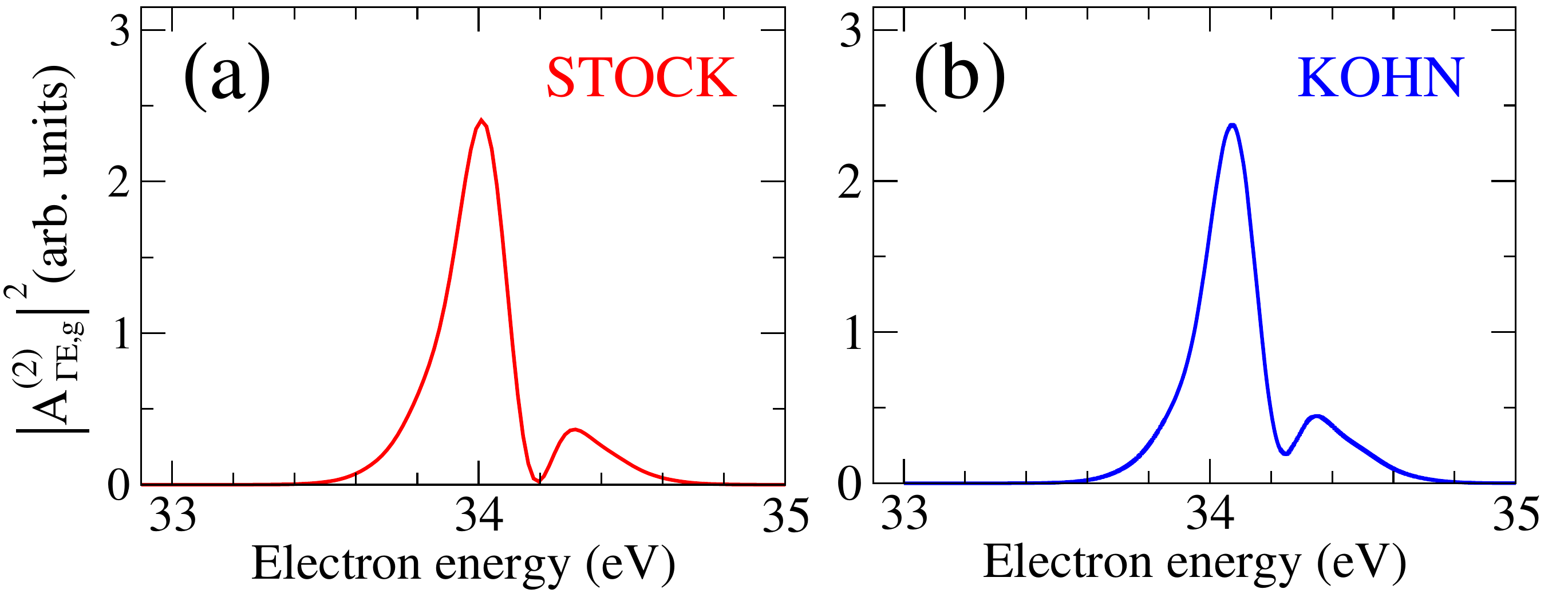}\vspace{-0.0 cm}\\
\includegraphics[width=8.25 cm]{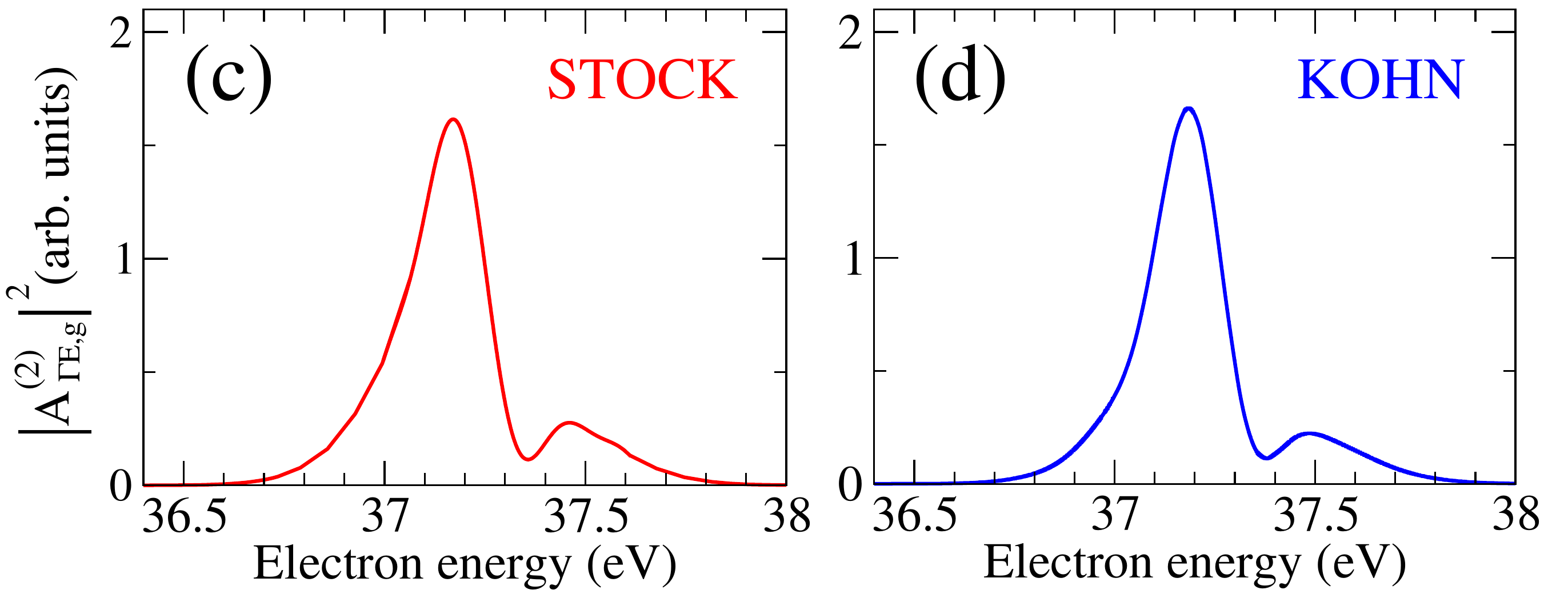}
\end{array}$
\caption{Same as Fig. \ref{fig:6} for an infrared pulse with wavelength $785$~nm and FWHM 26 fs, calculated at SB$_{-1}$ in (a) and (b), and 
at SB$_{1}$ in (c) and (d).}\label{fig:7}
\end{center}
\end{figure}

In the first set of calculations, we use an optical field with wavelength $\lambda=300$~nm, time delay $\tau=0$, intensity $2.5\times10^{11}$~W/cm$^2$, and two different pulse lengths with FWHM $30$~fs and $50$~fs. For such a small wavelength, the two-photon transition matrix element $\mathcal O_{\Gamma' E', \Gamma E}$ is sampled in a region with $E'-E\approx 4$~eV, which is very far from the diagonal band $E'=E$ where the effect of the regularization procedure is most visible. It should be noted, however, that the 2PTME in Eq. (\ref{eq:2PTME}) involves an integration over the complete energy interval, and hence the effect of the regularization procedure can not be entirely eliminated.

 \begin{figure}[t]
\begin{center}
$ \begin{array}{c}
\includegraphics[width=8.25 cm]{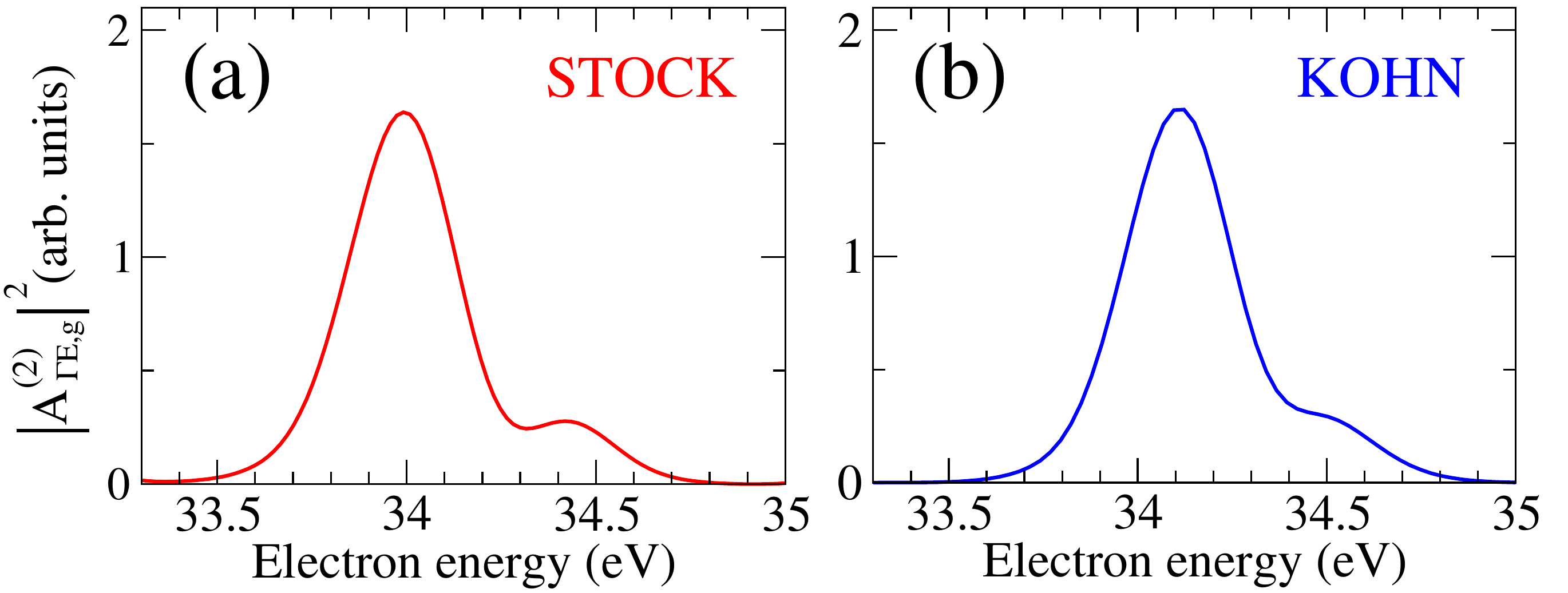}\vspace{-0.0 cm}\\
\includegraphics[width=8.25 cm]{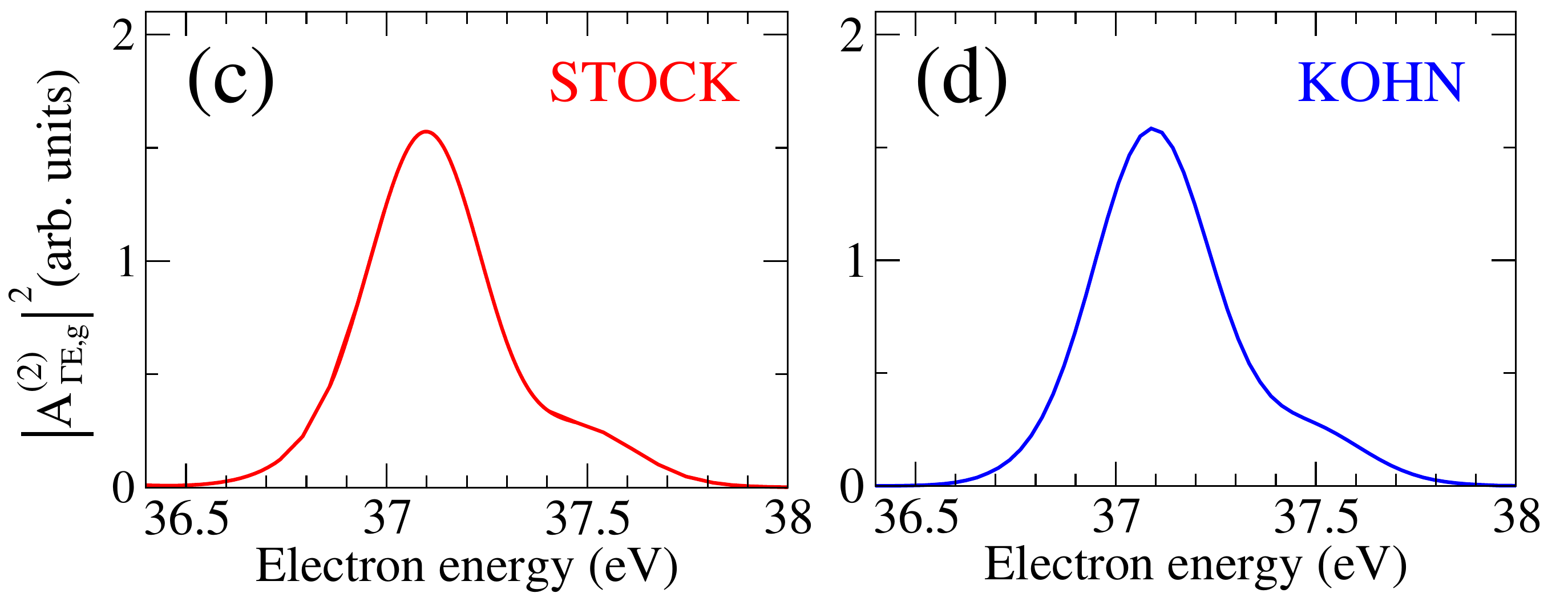}
\end{array}$
\caption{Same as Fig. \ref{fig:7} for an infrared pulse with wavelength $800$~nm and FWHM 13~fs.}\label{fig:8}
\end{center}
\end{figure}
Figure~\ref{fig:6} shows the partial ionization probability in {$^1$S} symmetry for both pulses at SB$_{-1}$. The \STOCK~and Kohn methods are in quite good agreement. In both cases, the resonant profile exhibits a small shoulder. As the pulse length is increased, in both calculations the shoulder becomes less pronounced, the main peak narrows, and it is accompanied by a flat pedestal. There are also some differences. In the Kohn method the shoulder is more pronounced and broader for the $30$~fs FWHM pulse, whereas the pedestal is smaller for the $50$~fs FWHM pulse.

\begin{figure*}[htbp]
\begin{center}
$ \begin{array}{ccc}
\includegraphics[width=5.5 cm]{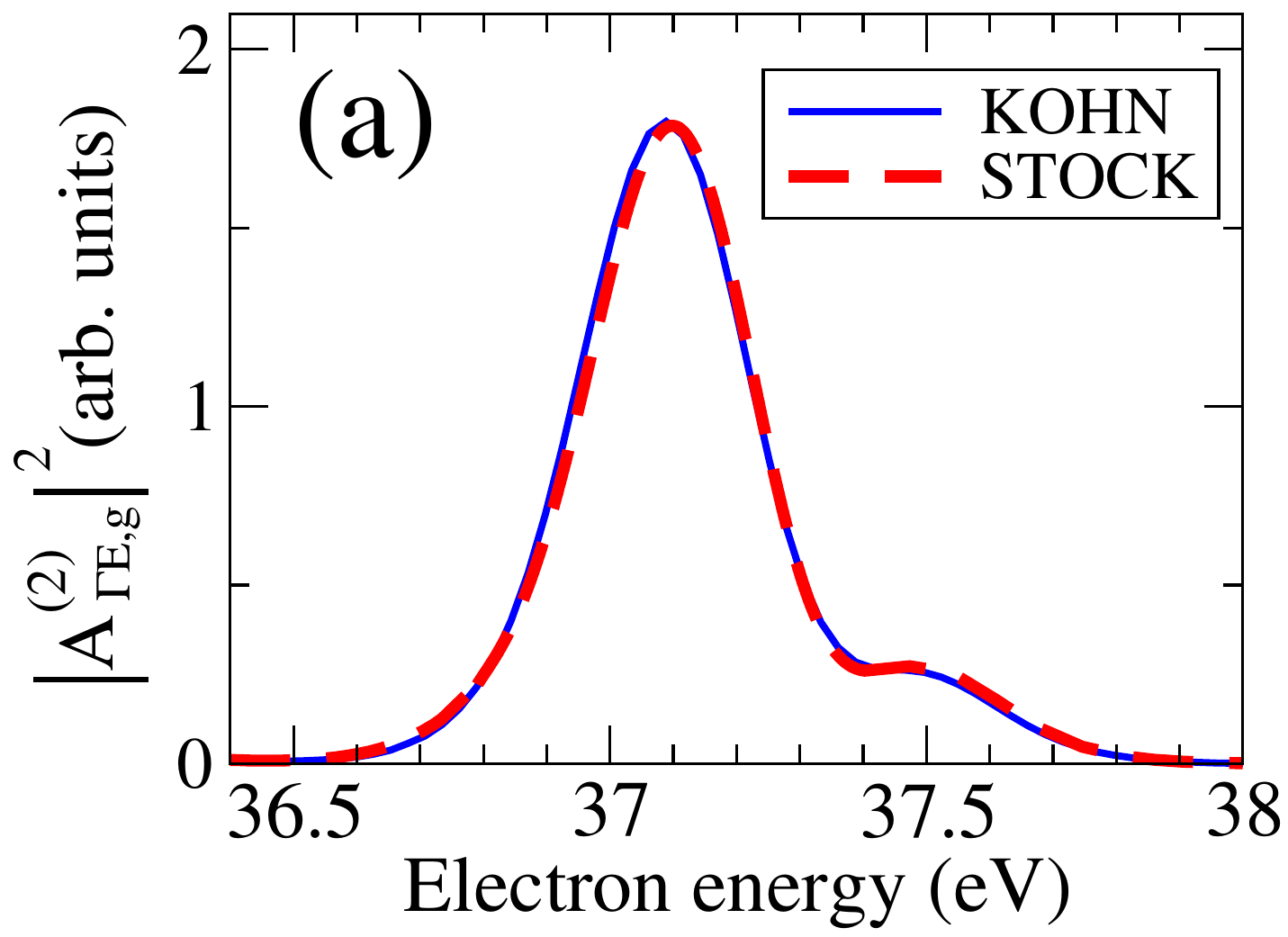} & \includegraphics[width=5.5 cm]{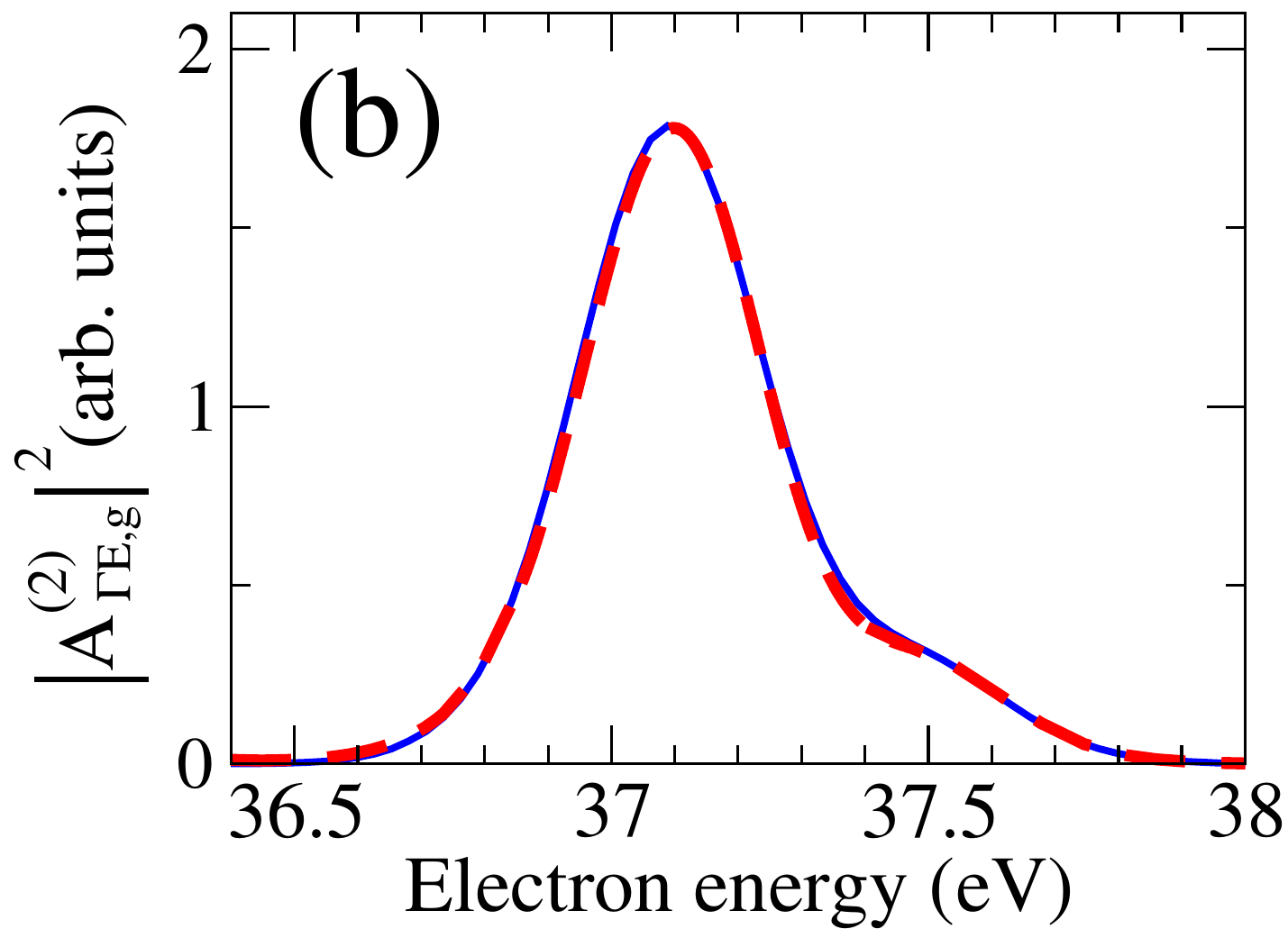} &\includegraphics[width=5.5 cm]{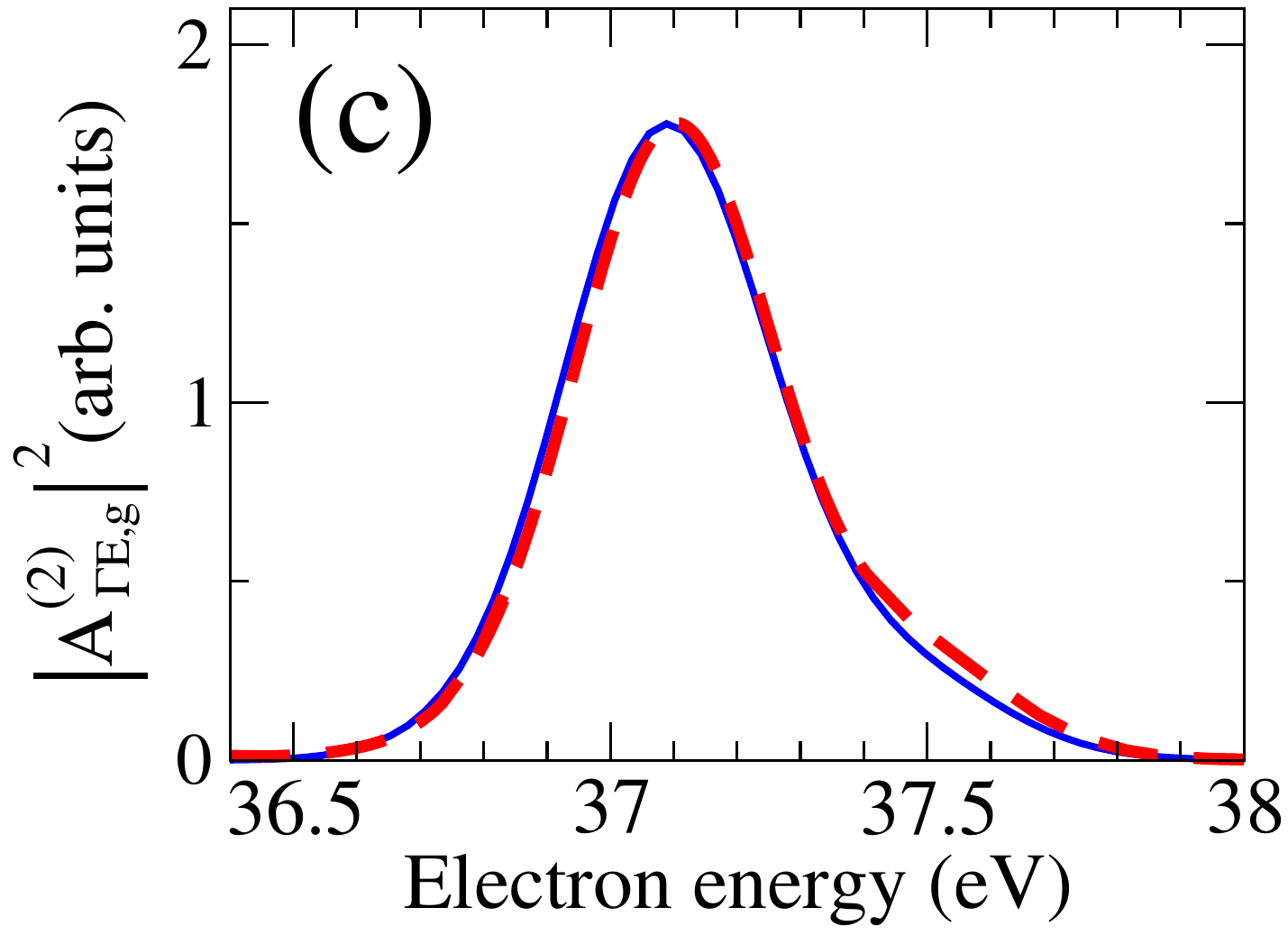}
\end{array}$
\caption{Same as Fig. \ref{fig:6} at the sideband SB$_{1}$ for an infrared pulse with wavelength $800$~nm and FWHM 13~fs, and for a time delay
 (a) $\tau=-1.45$~fs, (b) $\tau=0$~fs, and (c) $\tau=1.45$~fs.}\label{fig:9}
\end{center}
\end{figure*}
Next, we consider a probe field with a larger wavelength $\lambda=785$~nm, intensity $2\times10^{11}$~W/cm$^2$, and pulse duration, FWHM 26~fs. The results at both sidebands are shown in Fig.~\ref{fig:7}. Once again, the two methods are in good agreement, with a shoulder well delimited by a clear minimum.  The agreement is nearly perfect for the upper sideband, SB$_{1}$, while the value of the minimum at SB$_{-1}$ differs slightly in the two calculations.

Figure~\ref{fig:8} shows the result of a similar calculation conducted employing an IR pulse with a slightly longer wavelength, $\lambda=800$~nm, and significantly shorter duration, 13~fs FWHM.
As expected, the peaks are broader and the resonance modulation on the ionization probability is less pronounced than for the case of the pulse durations of FWHM 26~fs (see Fig. \ref{fig:7}). 

The results presented so far have been obtained setting the pump-probe delay to zero ($\tau=0$). This is because the two-photon scheme examined in Fig.~\ref{fig:6}(a) does not give rise to multiple interfering ionization pathways, and hence, the photoelectron spectrum does not exhibit any rapid sinusoidal modulation as a function of the time delay. 
Conversely, any change in the two-photon ionization probability as a function of $\tau$ is a sensitive probe of the accuracy of the Kohn calculation. Here, we use again a short pulse ($\lambda=800$~nm and FWHM 13~fs), and plot in Fig. \ref{fig:9} the $^1S$ ionization probability at SB$_1$, calculated with the two methods, at $\tau=-1.45, 0$, and $1.45$~fs. Note the value $\tau =1.45$~fs is only slightly larger than half a period of the IR but a significant change is seen compared to zero time delay. 
In this case, the two methods are in excellent agreement.

\begin{figure}[b]
\includegraphics[width=8.0cm]{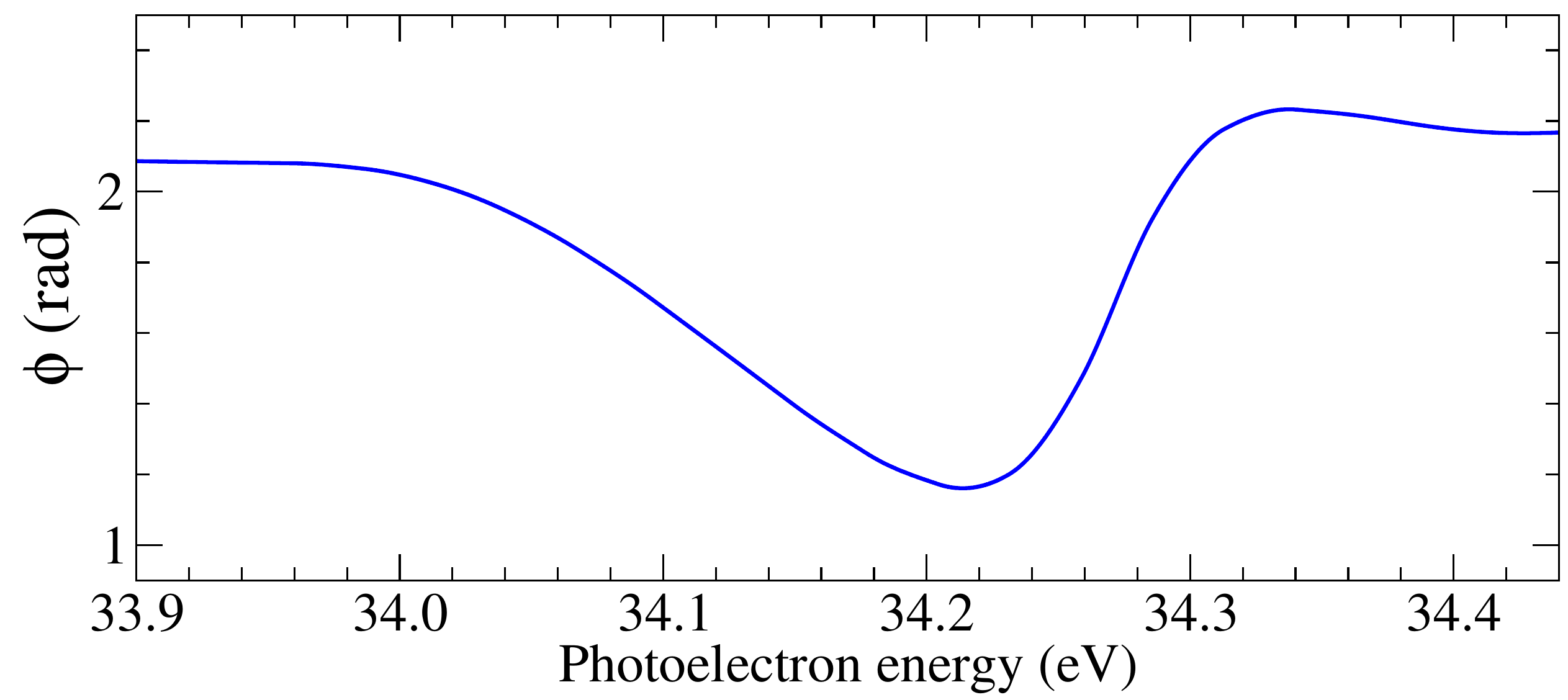}
\caption{Phase of the two-photon ionization amplitude in $^1S$ symmetry extracted using the RABBITT technique (see text for details).}
\label{fig:10}
\end{figure}
Finally, we consider the RABBITT scheme presented in \hbox{Fig. \ref{fig:5}(b)}. Here, we use again an IR pulse with wavelength  $800$~nm and FWHM 26~fs to limit envelope effects for time delays spanning a few IR periods around $\tau=0$.  We use two Gaussian XUV pulses: one with central frequency $\omega_0$ tuned at the resonance and which gives rise to the two-photon amplitude $\mathcal A^{(2)}_R$ at SB$_1$, and one with central frequency $\omega_1=\omega_0-2\Omega$, giving rise to a non-resonant two-photon amplitude $\mathcal A^{(2)}_{NR}$ at SB$_1$. The resulting total $^1S$ photoelectron amplitude is $\mathcal A^{(2)}_{\Gamma E,g}=\mathcal A^{(2)}_{R} + \mathcal A^{(2)}_{NR}$. The photoelectron spectrum is 
\begin{equation}
\left|\mathcal A^{(2)}_{\Gamma E,g}\right|^2=\left|\mathcal A^{(2)}_{R}\right|^2+\left|\mathcal A^{(2)}_{NR}\right|^2+2\left|\mathcal A^{(2)}_{R}\right|\left|\mathcal A^{(2)}_{NR}\right|\cos(2\Omega\tau+\phi),
\end{equation}
where
\begin{equation}
\label{eq:resonant_nonrenonsant}
\mathcal A^{(2)}_{R}=\left|\mathcal A^{(2)}_{R}\right|e^{i(\phi_R-\Omega\tau)}~~;~~\mathcal A^{(2)}_{NR}=\left|\mathcal A^{(2)}_{NR}\right|e^{i(\phi_{NR}+\Omega\tau)},
\end{equation}
with $\phi=\phi_R-\phi_{NR}$.  The phases $\phi_R$ and $\phi_{NR}$ are the resonant and nonresonant two-photon ionization phases, respectively.
To a good approximation, $\left|\mathcal A^{(2)}_{R}\right|$ and $\left|\mathcal A^{(2)}_{NR}\right|$ 
are independent of the time delay  in the neighborhood of $\tau=0$. As a result, the apparent phase shift of the $2\Omega$ oscillation of the signal coincides with $\phi$. Therefore, instead of fitting the signal as a function of $\tau$ to compute $\phi$, we can directly extract it from the \emph{ab initio} amplitudes computed at $\tau=0$ as
 \begin{equation}
 \phi=\arg\left[\mathcal A^{(2)}_{R}\mathcal A^{(2)*}_{NR}\right].
 \end{equation}
 The resulting $\phi$, shown in Fig. \ref{fig:10} as a function of the photoelectron energy, exhibits the characteristic sigmoidal resonant modulation observed in past experimental and finite-pulse theoretical studies~\cite{PhysRevA.93.023429,Gruson734,Kotur2016,PhysRevLett.113.263001,Busto18}. Using a longer pulse and a methodology similar to the rainbow RABBITT technique, would enable the extraction of the 2PTME using the Kohn method.
 
So far, our implementation of the two-photon Kohn method has shown a convincing agreement with the \STOCK~method, which supports the idea that it can represent a viable theoretical approach to the attosecond spectroscopy of molecules in perturbative regime. Yet minor differences with the benchmark and possible improvements deserve some comments. 
 
First, the results of one- and two-photon ionization in many-body systems on a hybrid basis are very sensitive to the level at which correlation is treated, as well as to the convergence of the radial basis in individual close-coupling channels. Whereas the description of He$^+$ states is essentially exact, and the He ground state is computed with comparable accuracy, in the two approaches, the scattering states in the \STOCK~method are less constrained and hence supposedly more accurate than those obtained with the relatively small-size calculation used in the Kohn method. 
 
Second, the finite-pulse calculations with the MESA+CK and the simulations with the STOCK method are computed in two  different gauges, which are expected to coincide only in the limit of a complete basis for any model Hamiltonian with a local potential. 

Third, the regularization procedure used in the Kohn method to compute free-free transition dipole moments in length form is expected to have some repercussion on the final results. This approximation can be improved by increasing the grid size. In the present implementation, however, the free-free integrals are calculated mapping the continuous functions on a three-dimensional grid. As a result, the size of the calculation increases with the cube of the linear size of the quantization box, which become rapidly too expensive as the maximum radius is increased. This limitation can be circumvented in future implementations, since at distances larger than the region where the Gaussian primitive functions have an appreciable value, the free-free integrals can be computed on a one-dimensional radial grid, which would drastically reduce the size of the calculation. Another possibility is to use analytical values between shifted Coulomb functions beyond a given boundary, regularized as prescribed in Refs.~\cite{Madajczyk89,Veniard90}. Finally, the Kohn method can be extended to calculate transition dipole moments in velocity form where only a simple pole exists and exact formula can be used. 
 
 \section{Conclusions and Outlook}
  \label{sec:3}
We have developed and presented a new method to compute atomic and molecular pump-probe photoelectron spectra in the perturbative regime, based on variational multi-electron continuum wave functions obtained from the complex variational Kohn method. We have used two-photon ionization of helium near the ($2s2p$)$^1P^o_1$ resonance as a proof of principles of our approach and compared it with results obtained using \STOCK, an atomic B-spline close-coupling package. The free-free transition dipole moment obtained in the two methods are in remarkable agreement. We have also compared the photoelectron spectra predicted by the two methods for various combinations of XUV and optical/IR pulses with different wavelength, length, and delay, again finding good agreement.

The new implementation of the Kohn method can efficiently compute one- and two-photon ionization amplitudes and can be applied to large molecules. It could be used to compute both integral and angle-resolved observables in pump-probe experiments, such as product yields in molecular dissociation. In principle, the perturbative approach can be extended to compute still higher-order multi-photon amplitudes, which would open the way to perturbative \emph{ab initio} estimate of attosecond transient absorption spectra in the weak-field regime.

\section*{Acknowledgments}
The authors would like to thank T.~N.~Rescigno for helpful discussions.
The work of N.D. and L.A.~was supported by the United States National 
Science Foundation under NSF grant No.~PHY-1607588 and the work of  B.I.S 
was supported by the Department of Commerce, 
National Institute of Standards and Technology.


\bibliography{Atto.bib}

\end{document}